\documentclass{aastex}
\usepackage{emulateapj5}
\usepackage{apjfonts}
\usepackage{epsf}
\usepackage{lscape}
\usepackage{timesfonts}

\def\cale{{\cal E}}
\def\ergs{$\rm erg~s^{-1}$}
\def\kms{\ifmmode ${\rm km~s^{-1}}$ \else ${\rm km~s^{-1}}$ \fi}
\def\oiii{\ifmmode [O {\sc iii}] \else [O {\sc iii}] \fi}
\def\feii{\ifmmode [Fe {\sc ii}] \else [Fe {\sc ii}] \fi}
\def\nh{\ifmmode N_{\rm H} \else N_{\rm H} \fi}
\def\mbh{M_{\rm BH}}
\def\sunm{M_{\odot}}

\journalinfo{The Astrophysical Journal, Vol. 661, 2007, June 1, in Press}

\begin{document}

\title{The Unified Model of Active Galactic Nuclei:  II. Evolutionary Connection}

\author{Jian-Min Wang and En-Peng Zhang}

\affil{Key Laboratory for Particle Astrophysics, Institute of High Energy Physics, CAS, Beijing 100049, China.}

\slugcomment{Received 2006 September 2; accepted 2007 January 30;}
\shorttitle{Unified Model of AGNs}
\shortauthors{WANG AND ZHANG}

\begin{abstract}
The classical unified model is challenged by the presence of various kinds of Seyfert 2 galaxies. We assemble a sample 
composed of 243 nearby Seyfert galaxies with redshifts $z\le 0.05$ to test the unification scheme. The sample contains 
94 broad line Seyfert 1s (BLS1s), 44 narrow line Seyfert 1s (NLS1s), 36 X-ray absorbed
hidden broad line region (BLR) Seyfert 2s
(HBLR S2s), 42 X-ray absorbed non-hidden BLR Seyfert 2s (non-HBLR S2s) and 27 X-ray unabsorbed Seyfert 2s 
(unabsorbed non-HBLR S2s and HBLR S2s). 
We examine the black hole mass and accretion rate distributions in the sample. We find that: 
1) NLS1s have less massive black hole masses than BLS1s; 2) HBLRS2s 
have the same mass distribution of the black holes with BLS1s; 3) the absorbed non-HBLR S2s  have less massive 
black holes than HBLR S2s; 4) unabsorbed non-HBLR S2s have the most massive black holes. We thus 
have a queue of black hole masses from small to large: narrow to broad line Seyfert galaxies, providing 
new evidence for the evolutionary sequence of Seyfert galaxies. We find that the opening angles of the torus 
in NLS1s and absorbed non-HBLR S2s are significantly smaller than that in BLS1s and HBLR S2s. The growth of 
the black holes and increases of the opening angles of the tori determine the various appearances of 
Seyfert galaxies. We also find that the unabsorbed Seyfert 2 galaxies could be caused by 
low gas-to-dust ratios in the present sample. This indicates that the star formation histories could be different 
in the unabsorbed from in absorbed Seyfert 2 galaxies, showing evidence for suppressed star formation by black hole
activities. We outline a new unification scheme based on the orientation hypothesis: Seyfert galaxies can be unified 
by including growth of black holes, Eddington ratios, changing opening angles of tori and gas-to-dust ratios in the 
tori. Seyfert galaxies are tending to finally evolve to unabsorbed 
non-HBLR Seyfert 2 galaxies, in which the black holes are accreting with low accretion rates and both the broad 
line region and dusty torus disappear. 

\end{abstract}

\section{Introduction}
Seyfert galaxies have received much attention during the last several decades from radio to hard X-rays. They 
were simply divided into two classes: Seyfert 1 and 2 with and without Balmer broad emission lines, respectively. 
This classification can be physically explained by obscuration of a dusty torus with different orientation with
respect to 
observers  based on a spectropolarimetric observation of NGC 1068 (Antonucci \& Miller 1985). Such a straw person 
model (SPM) of the unification scheme, is able to explain the main different characters of the two kind Seyfert 
galaxies (see a review of Antonucci 1993). The orientation-based unification scheme then becomes popular, but 
it also receives confronted evidence with more sophisticated observations. It has been suggested that there 
are several kinds of Seyfert 2 nuclei according to their diversities of properties of
emission lines and multiwave continuum behaviors, which are needed more physical parameters beside the
orientation (Tran 1995, 2001, 
2003). This indeed constitutes serious challenges to the traditional model of unification scheme.

We list {\em all} known types of Seyfert galaxies in 
Table 1 for their abbreviations and properties. There is increasing evidence for that only orientation is not able 
to unify the Seyfert galaxies with the presence of diversities of multiwave properties. 
Fig. 1 outlines possible relations among them. Seyfert 2 galaxies show much more 
complicate properties, such as contradictory properties in optical band and X-rays, and transition from type II to 
type I etc., but also NLS1s appear as a special group, containing less massive black holes and having higher 
Eddington ratios (Boller et al. 1996; Wang et al. 1999; Wang \& Zhou 1999; Mineshige et al. 2000; Wang \& Netzer 2003).
It is generally believed that BLS1s are the counterparts of absorbed HBLR S2s in light of the 
traditional unification scheme, but their central engines have not been tested so far. Zhang \& Wang (2006, hereafter
ZW06) suggest that the absorbed non-HBLR S2s could be counterparts of NLS1s, but viewed
at high orientation and only contain a "narrower" broad line region. 
Williams et al. (2002) reported an optically-selected NLS1 sample from SDSS that the black holes may have lower 
accretion rates, but still close to the Eddington limit. It could  be plausible that the optically-selected 
NLS1s just have increasing accretion rates and appear with a relatively flat spectrum in soft X-ray band. 
Both NLS1s and absorbed non-HBLR S2s have less massive black holes and relatively high accretion 
rates in the suggested scenario of the present paper. 
Tran (2003) recently showed that HBLR S2s are intrinsically more powerful than non-HBLR S2s from a 
large spectropolarimetric study of Seyfert 2 galaxies. X-ray observations show the presence of two kinds of Seyfert 2 
galaxies: weakly and strongly absorbed Seyfert 2s. What are relations among absorbed/unabsorbed HBLR 
S2s and non-HBLR Seyfert 2s? Except for orientation, physical parameters, such as
the opening angle (or clumpness), black hole mass and accretion rate, 
geometry and composition of the torus are expected to play an important role in the unification of the Seyfert zoo. 

\figurenum{1}
\begin{figure*}[t]
\centerline{\includegraphics[angle=-90,width=14.5cm]{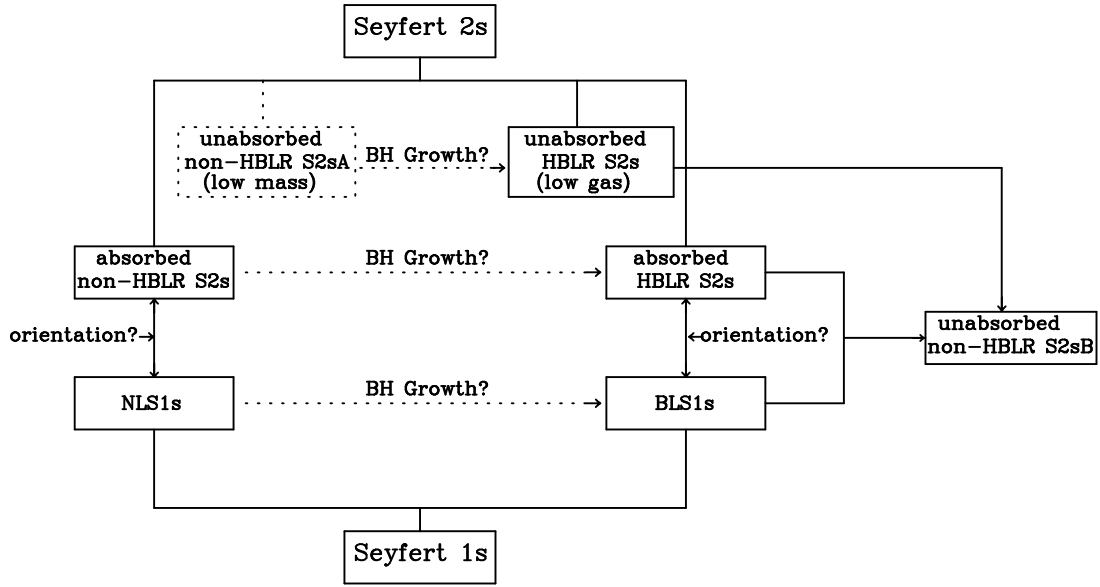}}
\figcaption{The Seyfert Zoo and their possible relations of sub-class Seyfert galaxies suggested in this paper.
The abbreviations can be found in Table 1. The absorbed HBLR S2s and BLS1s can be unified by obscuration of a 
dusty torus (Antonucci \& Miller 1985) whereas the absorbed non-HBLR S2s and NLS1s can be done in the same way 
(Zhang \& Wang 2006). Orientations and opening angles of the dusty tori play 
key roles in unifications of the absorbed non-HLBR S2s and NLS1s, HBLR S2s and BLS1s, but the growth of the black 
holes, the accretion rates and gas-to-dust ratios should be included for the unification scheme of Seyfert galaxies. 
This plot can be converted into an evolutionary sequence of Seyfert galaxies (see Figure 17). 
} 
\label{fig1}
\end{figure*}

The geometry and structure of torus remain a matter of controversy. The torus could be clumpy or mixed with 
star formation region (Elitzur 2005) and the reflecting mirror of the BLR has complicate structure and ingredients
(hot electrons and various size of dust grains) (Smith et al. 2005). This allows appearances of various kinds of 
Seyfert 2 galaxies (see Tab 1 and Fig. 1), which makes 
the classification and unification complicate (Panessa \& Bassini 2002, hereafter PB02; Tran 2003; ZW06; Gallo 
et al. 2006, hereafter G06). A fast transition within 6 weeks from Compton-thick to Compton-thin in a 
Seyfert 1.8 galaxy NGC 1365 implies that there are random motions of dusty clouds inside the torus (Risaliti et 
al. 2005). Infrared detection of dusty cloud distribution shows clear structure in 20-50pc around the central 
engine in NGC 1068 (Tomono et al. 2006). Whether the dusty torus can be described
in a unified form is still open. 

On the other hand, the opening angle of the dusty torus is an important ingredient in the unification scheme.
There is a strong evidence for an evolution of the opening angle of the torus with hard X-ray luminosity, 
displaying a receding torus (Hasinger 2004; Steffen et al. 2004; Wang et al. 2005; Barger et al. 2005). 
Hard X-ray surveys by {\em Chandra} and {\em ASCA} have revealed that the ratio (${\cal R}$) of type II to I AGNs 
is decreasing with the hard X-ray luminosity ($L_{\rm HX}$) (Ueda et al. 2003; Steffen et al. 2003; Hasinger 2004). 
This result has been confirmed by {\em Hubble Space Telescope} identifications of the narrow emission lines (Barger 
et al. 2005). This strongly indicates that the opening angle of the dusty torus is changing rather than a constant. 
Wang et al. (2005) show that the covering factor ${\cal C}$ of the torus, estimated by the ratio of infrared 
to optical-UV luminosity, also decreases with the hard X-ray luminosity in PG quasars and the ${\cal C}-L_{\rm HX}$ 
relation is consistent with ${\cal R}-L_{\rm HX}$. This directly provides obvious evidence for the 
consistencies among the obscuring material of BLR, reprocessing and absorbing 
material of X-rays. The appearance of type II AGNs is not only due to the orientation of the
dusty torus, but also strongly depends on its opening angle. Clearly, the unification
scheme should include the evolution of the opening angle of the torus.

The composition of the torus also makes the appearance of Seyfert galaxies complicate. The X-ray weak absorption 
in an optically-defined Seyfert 2 galaxies indicates there is low absorption material along observer's line. 
Panessa \& Bassani (2002, hereafter PB02) assembled a sample composed of 17 unabsorbed Seyfert 2 galaxies with 
low X-ray absorption and suggested that the black holes in unabsorbed S2s have very low accretion rates
since most of objects in this sample have relatively low optical luminosity. Seyfert 2 galaxies could be "true" 
Seyfert 2 objects due to the absence of a broad line region provided the black holes have accretion rate lower than 
a critical one (Laor 2003; Nicastro et al. 2003). On the other hand, G06 find about $50\%$ {\em ROSAT}-selected AGNs 
are Seyfert 2 galaxies and Wolter et al. (2005) find three type II quasars only have low absorption, but their optical 
properties are not abnormal. The gas-to-dust ratio (${\cal Q}$) could be an important parameter in determining the X-ray 
properties of an optically-defined Seyfert 2 galaxy (Maiolino et al. 2001a,b; Boller et al. 2003; Galo et al. 2006). 
It seems to us that there are, at least, two kinds of unabsorbed Seyfert 2 galaxies.
PB02 and G06 samples likely represent the two kinds, which are low and normal optical luminosity AGNs, respectively. 
Additionally, optically-defined Seyfert 2 galaxies, F 01475-0740 and NGC 2992 
in Table 5, show polarized broad lines, but they have low absorption in X-rays (i.e. showing type I property). This 
indeed shows they have hidden 
broad line region, but why do they show type I properties in X-rays? The unabsorbed type II quasars are found to
have black holes with mass $>10^{10}\sunm$ and low Eddington ratios (Wolter et al. 2005). 
These evidence shows not only the Eddington ratio, but also the gas-to-dust ratio might play a key role in the
appearance of Seyfert 2 galaxies. Star formation/burst inside the torus (Collin et al. 1999) definitely produces amount 
of dust, does the gas-to-dust ratio determine the complicate appearance? and are the histories 
of the star formation rates different in the unabsorbed from the absorbed Seyfert 2 galaxies? 

The growth of the black holes is very significant during the activity cycle (Wang et al. 2006). For a black hole with 
a mass of $5\times 10^6\sunm$, it will grow up to $\sim 5\times 10^7\sunm$ within one fourth of the Salpeter time (see 
\S6.3) if the black hole accretes at the Eddington limit. For NLS1s and absorbed non-HBLR S2s, the growth timescale of 
the black holes is much shorter than one fourth of Salpeter time if they have super-Eddington rates (Kawaguchi et al. 
2004). With the BH growth, a narrow line Seyfert galaxy becomes broad line one if it still has high enough accretion 
rates. It can thus not be avoided to include the black hole growth in the unification scheme. An obvious connection 
comes out reasonably: the inevitable growth of black holes due to accretion is driving evolution of less massive 
black holes into broad line Seyfert galaxies. 

The increasing evidence strongly suggests that the canonic unification scheme should be improved by supplementing 
three possible ingredients: 1) black hole masses; 2) opening angles of tori; 3) gas-to-dust ratios in the tori.
We outline a possible evolutionary and orientation connection for Seyfert galaxies in Figure 1 through issuing
the following main questions:

1. Are there evolutionary sequences of the black hole masses and accretion rates among Seyfert galaxies?

2. Are there sequences of the opening angles from narrow to broad Seyfert galaxies?

Figure 1 suggests the unabsorbed non-HBLR Seyfert 2 galaxies
could be the final stage of {\em all} other types when the torus is going to be depleted and accretion rates 
are thus very low. If this scenario is correct, the phenomena of Seyfert galaxies could be understood profoundly.

The present paper is devoted to test this evolutionary idea outlined in Figure 1. We give the sample in \S2.
In \S3, we estimate the black hole masses and accretion rates of BLS1s and absorbed HBLR S2s. Unabsorbed Seyfert 2s
are discussed in \S4. We discuss starburst in circumnuclear regions in \S5.
We devote \S6 to discuss the evolution of central engines of Seyfert galaxies. More extensive 
discussions are given in \S7. Finally, we draw a conclusion that the orientation-based unification model 
should be improved by inclusion of the evolutions of the black holes, torus opening angles and gas-to-dust ratios.
Seyfert galaxies will be unified within such a scenario.

\section{Sample and Data}
For the main goals of the present paper, we have to define an homogeneous and complete Seyfert galaxy sample. However 
this is not an easy job because of  diversities of their multiwavelength properties. There are several constraints 
on sample selection mainly arisen from: 1) polarization identifications of non-HBLR S2s; 2) absence of X-ray 
observations; 3) appearance frequencies of NLS1s survey in different bands. The classical Seyfert 1 galaxies 
are extensively well-studied, however, only some Seyfert 2 galaxies have been observed by spectropolarmetry 
since this is a time-consuming and tedious job (Tran 2003). The strongest constraint 
on the present sample is due to non-HBLR S2s. We have searched all the papers related with Seyfert galaxy 
samples systematically studied and made a combined sample. Table 2 lists the references since 1985. 

Figure 2. shows the redshift distributions are inhomogeneous in the all objects taken from the references listed 
in Table 2. We find
most of non-HBLR S2s are limited within $z\le 0.05$. We thus limits the present sample within $z\le 0.05$ to 
avoid the luminosity-biased effects, but this does {\em not} mean the combined sample is volume limited in the 
present paper. Hao et al. (2005) present the \oiii luminosity function of very low luminosity AGNs 
($L_{\rm [O III]}=10^5\sim 10^8L_{\odot}$), but its statistic properties of emission lines and continuum are not 
currently known. This makes it impossible to compare the present sample with Hao et

\figurenum{2}
\centerline{\includegraphics[angle=-90,width=8.0cm]{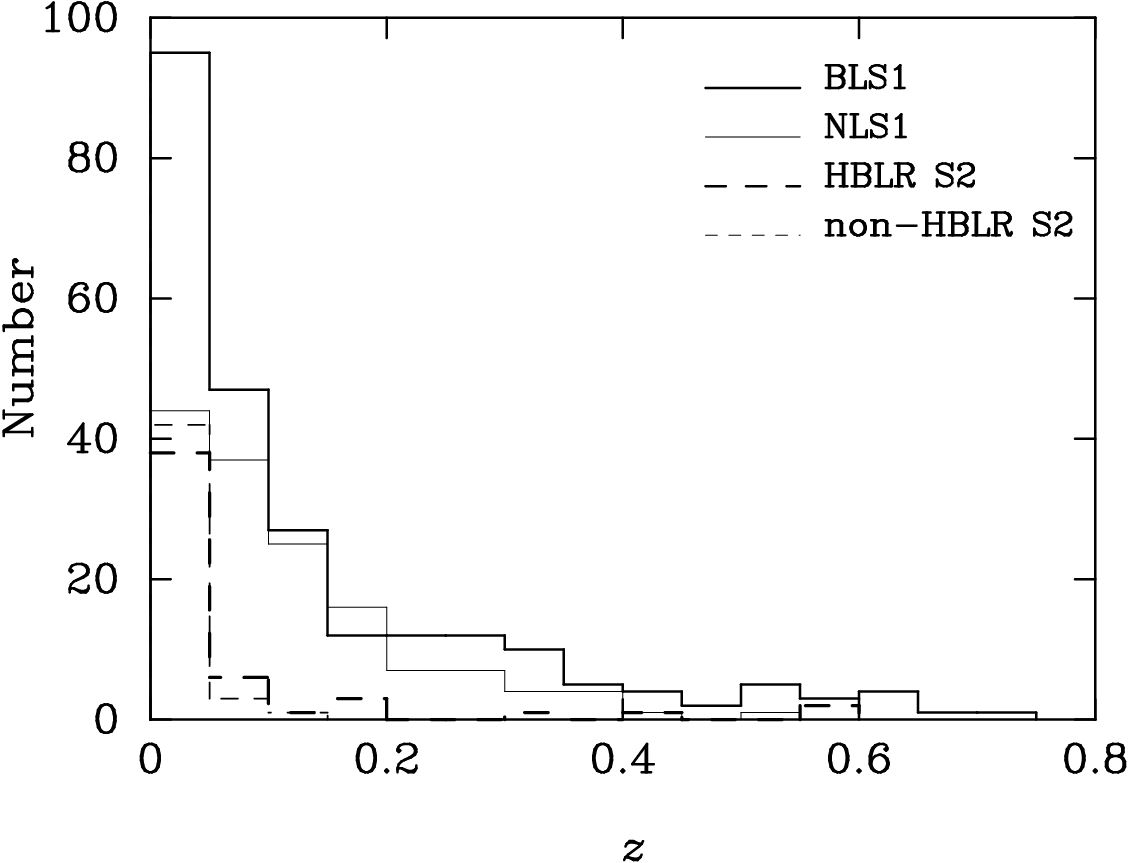}}
\figcaption{The redshift distribution of Seyfert galaxies from the references listed in Tab. 2. and Tab. 1-2 in ZW06.}
\label{fig2} 
\vglue 0.2cm

\noindent al's sample. A relatively 
complete sample is 12$\mu$m sample for nuclei brighter than 12$\mu$ luminosity $L_{\rm 12\mu}=10^{42}$\ergs 
with a mean $\langle V/V_{\rm max}\rangle\approx 0.46$ (Rush et al. 1993). However, this is a 
near IR-selected sample, in which NLS1's fraction is much smaller than that in 
{\em ROSAT}-sample (Grupe et al. 1998). We collect {\em all} NLS1s from the literatures, which  
are mainly from Boller et al. (1996), Grupe et al. (1998), Veron-Cetty et al. (2001), Stepain et al. (2003) 
and Williams et al. (2002). With the redshift limit, ZW06 gave the NLS1 
and absorbed non-HBLR S2 sample in their Table 2 and 3. BLS1s are mainly from Marziani et al. (1998) etc. 
and HBLR S2s from Maiolino \& Rieke (1995). Tables 3  and 4 give BLS1s and  
absorbed HBLR S2s, respectively. The weakly absorbed or unabsorbed S2s are mainly 
from Panessa \& Bassani (2002) and Gallo et al. (2006) and given in Table 5. The unabsorbed Seyfert 2s are divided 
into two sub-classes: unabsorbed non-HBLR Seyfert 2 and HBLR Seyfert 2 galaxies.
Therefore our sample covers: 1) BLS1s; 2) NLS1s; 3) X-ray absorbed HBLR S2s; 4) X-ray absorbed non-HBLR S2s; 5) X-ray
unabsorbed HBLR S2s; 6) X-ray unabsorbed non-HBLR S2s. We cover six kinds of Seyfert galaxies 
in this paper, but we realize that there could be a possibility dividing X-ray unabsorbed non-HBLR S2s into two
subclasses A and B, given in Table 1, which have different orders of black hole masses. 
Only 7 unabsorbed Seyfert 2 galaxies in Table 5 have been observed by spectropolarimetric 
observations, it is hard to judge whether they have hidden BLR or not without observations. However we could 
distinct them according to the theory recently developed\footnote{This artificial distinction makes it difficult
to discuss the distributions of the Eddington ratios, but distributions of BH masses. There are only seven
unabsorbed Seyfert 2s with spectropolarimetric observations. Additionally, there is
still possibility that the "non-HBLR S2s" are only upper limits to the BLR emission.}. 
There is a critical value of the Eddington ratio 
$10^{-3}$ based on bolometric luminosity $L_{\rm Bol}=10L_{2-10\rm keV}$ (Nicastro et al. 2003; Laor 2003), below 
which there is no hidden BLR (Nicastro et al. 2003). However, Panessa et al. (2006) used 
$L_{\rm Bol}=30L_{2-10\rm keV}$ to estimate the bolometric luminosity since $L_{\rm Bol}=10L_{2-10\rm keV}$ is 
underestimated. 
We thus take $\dot{m}_{\rm c1}=10^{-2.5}$ as the critical Eddington ratio to distinguish unabsorbed HBLR and 
non-HBLR Seyfert 2 galaxies. When the Eddington ratio is high enough ($\dot{m}_{\rm c2}\ge 0.2-3$), the broad line 
also disappears since the transition region from gas-pressure 
dominated to radiation pressure dominated moves outward (Nicastro 2000). We use the two criteria to classify the 
unabsorbed Seyfert 2 galaxies into non-HBLR and HBLR Seyfert 2 galaxies in Table 5. Finally we have total of 243 
Seyfert galaxies including 94 BLS1s, 44 NLS1s, 36 HBLR S2s, 42 absorbed non-HBLR S2s and 27 unabsorbed Seyfert 2s.

We compare the fractions of each kinds of Seyfert galaxies with 12$\mu$m sample (Rush et al. 1998) and 12$\mu$+CfA 
sample (Tran 2003) to display 
the completeness and homogeneity of the present large sample. Table 6 gives the comparisons. NLS1s are relatively 
poor objects in 12$\mu$m sample and Tran (2003) sample whereas
our sample covers more NLS1s because of including {\em ROSAT} X-ray survey results. The fractions of other sub-class 
Seyfert galaxies in our sample are just between the 12$\mu$m and Tran (2003) sample as shown in Table 6. The present 
sample is close to the homogeneity and completeness of Rush et al. (1993) and Tran (2003). We stress here that
the completeness of the low luminosity AGNs in the present sample is unknown, but the completeness of this sample
should be good enough compared with 12$\mu$ sample (brighter than $L_{12\mu}>10^{42}$\ergs).
We always compare the distributions of black hole masses and accretion rates
in the present sample with 12$\mu$ sample if possible. We pay much cautions when we draw conclusions related with 
them. Though the data are
from different authors and instruments, the qualities of the well-defined observables, for example, the
FWHM of H$\beta$, \oiii and continuum do not rely on the instruments much. 
Hence the statistical properties of Seyfert galaxies in this sample should be reliable. We adopt the following 
cosmological constant $H_0=75~\rm km~s^{-1}~Mpc^{-1}$, $\Omega_{\Lambda}=0.7$ and $\Omega_{\rm M}=0.3$ 
throughout the paper.

\section{BLS1s and Absorbed Hidden BLR S2s}
It has been realized that the width of hydrogen lines is dependent on the black hole masses and accretion rates
(Laor 2003; Nicastro et al. 2003). The basic idea of the unified model is that the central engines are same, 
but oriented at different viewing angles (Antonucci 1993). If the orientation-based unified model works, the 
black hole masses and accretion rates should be similar or same in BLS1s and HBLR S2s.
We first investigate this issue for the cases of BLS1s and HBLR S2s.

\subsection{Black hole mass}
There are several independent methods to estimate the black hole masses in type I AGNs, but we use
1) the values of black hole masses from the reverberation mapping if the objects have been measured (Kaspi et al.
2000; Peterson et al. 2004); 2) the empirical relation to estimate the black hole mass (Vestergaard et al. 2002,
Kaspi et al. 2005); or 3) the $M_{\rm BH}-\sigma$ relation, 
$\mbh=1.35\times 10^8\sunm\left(\sigma/200{\rm km~s^{-1}}\right)^{4.02}$ (Tremaine et al. 2002), where $\sigma$ is
the stellar velocity dispersion. It is suggested to replace $\sigma$ by FWHM([O {\sc iii}])/2.35 (Nelson \& Whittle 1995; 
Boroson 2003; Shields et al. 2003; Greene \& Ho 2005). This relation is based on evidence that the \oiii~ line 
width in AGNs is dominated by 
the gravitational potential on the scale of the host galaxy bulge (Nelson \& Whittle 1996; Nelson 2000).
We use the empirical reverberation relation (Vestgaard 2002) for Seyfert galaxies,
\begin{equation}
R_{\rm BLR}=32.9\left(\frac{L_{5100}}{10^{44}{\rm erg~s^{-1}}}\right)^{0.7}~{\rm lt-days}, 
\end{equation}
and the black hole mass can be obtained from 
\begin{equation}
\mbh=1.5\times 10^5\left(\frac{R_{\rm BLR}}{{\rm lt-days}}\right)
\left[\frac{{\rm FWHM(H\beta)}}{10^3{\rm km~s^{-1}}}\right]^2\sunm, 
\end{equation}
where we use $v=\sqrt{3}/2~{\rm FWHM}({\rm H\beta})$, $L_{5100}$ is the luminosity at 5100\AA~ and $R_{\rm BLR}$ 
is the BLR size.

\figurenum{3}
\centerline{\includegraphics[angle=-90,width=8.0cm]{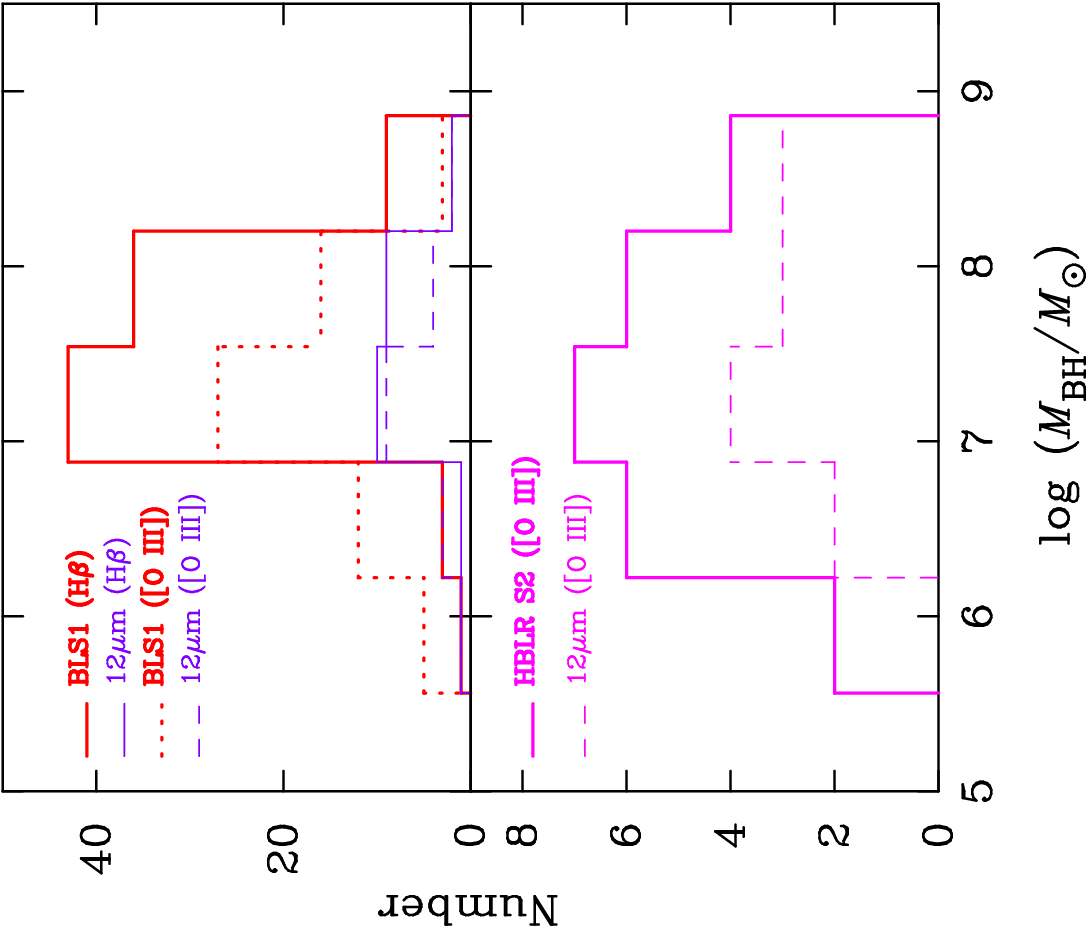}}
\figcaption{The distributions of the black hole masses in BLS1s and HBLR S2s. The different colors and line styles
represent Seyfert types and samples. We will not explain unless a new color and style line appear in the plot.}
\label{fig3}
\vglue 0.2cm

However, for Seyfert 2 galaxies, the  empirical reverberation relation is not available since the optical
continuum and broad emission lines are strongly absorbed by the dusty torus. 
For HBLR S2s and absorbed non-HBLR S2s, we use $\sigma={\rm FWHM}$([O {\sc iii}])/2.35/1.34 (Greene \& Ho 
2005) to obtain the black hole mass from the $\mbh-\sigma$ relation (Tremaine et al. 2002).
For some unabsorbed non-HBLR S2s, the \oiii luminosity is too faint to estimate BH mass from FWHM([O {\sc iii}]). 
Then we search the dispersion velocity of the galaxies from published literatures or
Lyon-Meudon Extragalactic Data Archive (LEDA), and estimate the masses of  non-HBLR S2s by the 
$\mbh-\sigma$ relation.   

Figure 3 shows distributions of black hole masses in BLS1s and HBLR S2s. First the distributions in the present 
sample are quite similar to that in the 12$\mu$m sample. This indicates the homogeneity and completeness of the 
present sample are similar to the 12$\mu$m sample, again. Second, the BH mass distributions obtained from [O {\sc iii}] 
width are very similar between BLS1s and HBLR S2s at a level of $p_{\rm null}=84.9\%$\footnote{The null probability 
$p_{\rm null}$ refers to that the two distributions are drawn at random from the same parent population.}. 
We note that the BH mass distributions 
in BLS1s estimated from \oiii width are different at low side from that by the reverberation empirical relation.
The average BH mass is $\langle \mbh\rangle=10^{7.53}\sunm$ in BLS1s. 
However, we should not directly compare the BH mass from different methods since these methods have their individual
scatters. On the other hand, the averaged black hole masses are 
$\langle \mbh\rangle=10^{7.24}\sunm$ and $\langle \mbh\rangle=10^{7.31}\sunm$ for BLS1s and HBLR S2s, respectively, 
and the peak values of the BH mass distributions are same also. This similarity can be also indicated by their host 
bulges (see below). Clearly the orientation-based unification gets supports from this similarity.

\figurenum{4}
\centerline{\includegraphics[angle=-90,width=8.0cm]{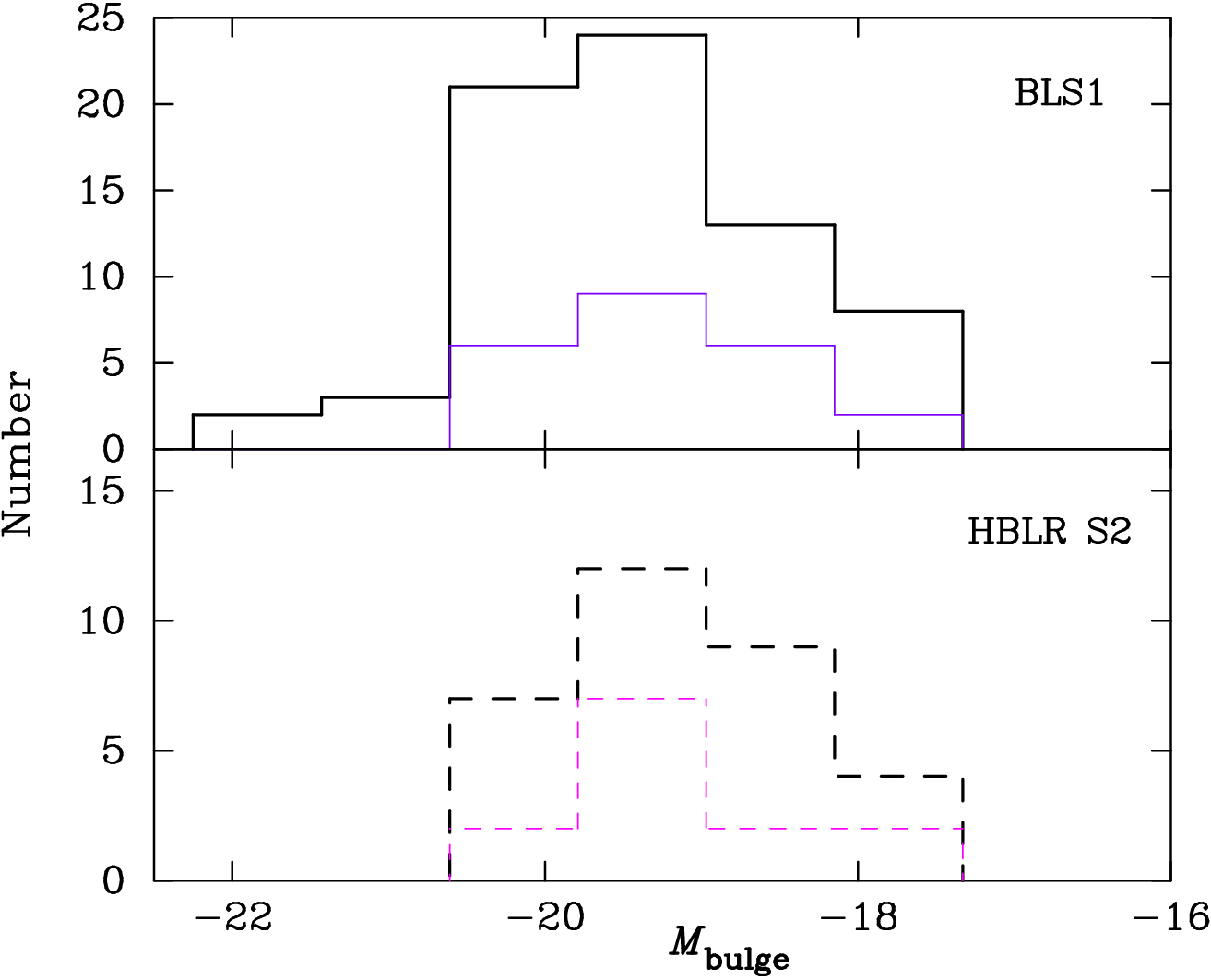}}
\figcaption{The distribution of the bulge luminosities in BLS1s and hidden broad line Seyfert 2
galaxies. The purple and red lines represent the 12$\mu$m sample.
The distributions of bulge luminosities in BLS1s and HBL S2s are similar to the 12$\mu$m sample.}
\label{fig4}
\vglue 0.2cm

To further test the black hole masses in Seyfert galaxies, we calculate the bulge luminosity from
Hubble stage $T$ and the total magnitudes ($M_{\rm tot}$) (Simien \& de Vaucouleurs 1986)
\begin{equation}
M_{\rm Bulge}=M_{\rm tot}-0.324\tau+0.054\tau^2-0.0047\tau^3
\end{equation}
where $\tau=T+5$. The distributions of bulge magnitudes are shown in Figure 4 for the present sample. We find that 
magnitude distributions of their bulges are very similar at a level of $p_{\rm null}=11.6\%$ and both 
of them are similar to 12$\mu$m sample. The average bulge magnitudes are $\langle M_{\rm Bulge}\rangle=-19.42$ and 
$\langle M_{\rm Bulge}\rangle=-18.97$ for BLS1s and HBLR S2s, respectively. The mean values of black holes masses 
are $\sim10^7\sunm$ for BLS1s and HBLR S2s. We note 
that there is difference at the bright side of bulge luminosity distribution. Except for the luminosity brighter 
than -20.5, the similarity is at a level of $15\%$. This result clearly indicates that the black hole masses 
in BLS1s and HBLR S2s should be similar based on the $\mbh-M_{\rm Bulge}$ relation (McLure \& Dunlop 2001;
H\"aring \& Rix 2004).

The similarities of the black hole mass distributions strongly favor the unification scheme. 
Being one of the main parameters of the central engines, $\mbh$ potentially indicates the evolutionary phase of 
the galaxies if the relation holds between the black holes and their host bulges. As we show below the accretion 
rates are also similar. This enhances the unification scheme based on orientation.

There are some uncertainties of the above estimations of the black hole masses. 
For the BLS1s, the empirical reverberation relation gives an uncertainty of  0.5 dex (Vestergaard 2002). 
The $M_{\rm BH} - \sigma$ relation has an intrinsic 
dispersion in $M_{\rm BH}$ that is about $\sim 0.3$dex (Tremaine et al. 2002).
The \oiii~ width can predict the black hole mass with a scatter of a factor 5 (Boroson 2003). For the
present HBLR S2 sample, we take the uncertainties to be 0.7dex. With these uncertainties,
we draw a conclusion that the BH mass distributions are similar in BLS1s and HBLR S2s for the present sample.

\subsection{Accretion rates}
\subsubsection{The Eddington ratios}
The accretion rates in these Seyfert galaxies can be estimated based on: 1) continuum of Seyfert 1 galaxies;
2) the \oiii luminosities. For Seyfert 1 galaxies, we take the popular formula,
\begin{equation}
L_{\rm Bol}\approx {\cal C}_{\rm B} L_{5100},
\end{equation}
where ${\cal C}_{\rm B}$ is the correction factor of the bolometric luminosity, ${\cal C}_{\rm B}=9$ (Kaspi 
et al. 2000), or ${\cal C}_{\rm B}=5$ (Netzer 2003). We use ${\cal C}_{\rm B}=9$ in this paper.
Since Seyfert 2 galaxies are basically obscured by the dusty tori, the only way to estimate the bolometric 
luminosity is from \oiii luminosity. Since the obscuration of the host galaxies  or extinction of the narrow 
line region, we have to correct \oiii flux. The \oiii fluxes are corrected for  BLS1s and HBLR S2s by the 
relation (Bassani et al. 1999)
\begin{equation}
F^{\rm cor}_{\rm [O III]}=F^{\rm obs}_{\rm [O III]}\left[\frac
{(\rm H \alpha /\rm H \beta)_{obs}}{(\rm H \alpha /\rm H \beta)_{0}}\right]
^{2.94},
\end{equation} 
assuming an intrinsic Balmer decrement $(\rm H\alpha /\rm H \beta)_{0}=3.0$ and $F_{\rm [OIII]}^{\rm obs}$ is
the observed \oiii flux. 
Here $(\rm H \alpha /\rm H \beta)_{obs}$ refers to the narrow components of H$\alpha$ and H$\beta$.
The bolometric luminosity is then estimated by
\begin{equation}
L_{\rm Bol}\approx 3500 L_{\rm [O III]},
\end{equation}
for both Seyfert 1 and 2s (Heckman et al. 2004),
where $L_{\rm [OIII]}$ is the corrected \oiii luminosity. This relation is based on the assumptions that the \oiii line 
is isotropically photoionized by the
central engine and  the covering factor of the narrow line region is a constant. The corrected luminosities 
are given in Table 3-5. With some uncertainties, it is a good indicator of ionizing sources in 
the first approximation.

\figurenum{5}
\centerline{\includegraphics[angle=-90,width=8.0cm]{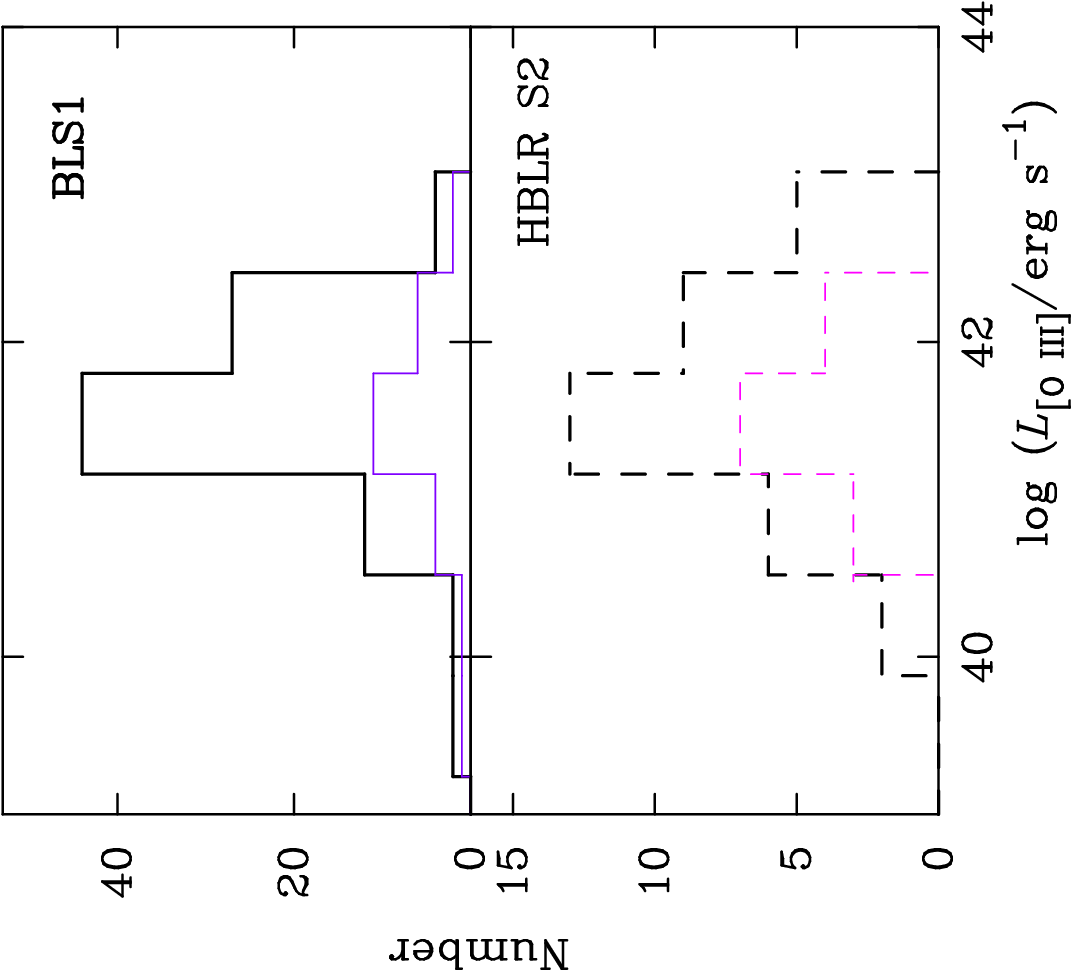}}
\figcaption{The distribution of the \oiii luminosities in BLS1s and hidden broad line Seyfert 2
galaxies. The purple and red lines represent the 12$\mu$m sample.
The \oiii luminosity distributions are similar to the 12$\mu$m sample.
BLS1s and HBLS1s have similar \oiii luminosity distributions, supporting the unification scheme.}
\label{fig5}
\vglue 0.2cm

Fig. 5 shows the distributions of \oiii luminosities. We find that they are similar at a level of
$p_{\rm null}=16.7\%$. The average \oiii luminosities are $\langle L_{\rm [O III]}\rangle=10^{41.57}$\ergs
and $\langle L_{\rm [O III]}\rangle=10^{41.67}$\ergs for BLS1s and HBLR S2s, respectively.
The bolometric luminosities are thus similar in BLS1s and HBLR S2s. If the black holes in the 
two kinds of Seyfert galaxies have the same radiative efficiency ($\eta=0.1$), the accretion rates 
$\dot{M}=L_{\rm Bol}/\eta c^2$ should be same. 

The importance of the Eddington ratio of the central engines has been extensively realized, for
example, it has influence on the hot corona of accretion disks (Wang et al. 2004), metallicity (Shemmer 
et al. 2004) and C {\sc iv} line emission (Baskin \& Laor 2005). With the bolometric luminosity, we can 
easily get the Eddington ratio
\begin{equation}
\cale=\frac{L_{\rm Bol}}{L_{\rm Edd}}
     =0.1\left(\frac{L_{\rm Bol}}{1.4\times 10^{44}{\rm erg~s^{-1}}}\right)
          \left(\frac{\mbh}{10^7\sunm}\right)^{-1},
\end{equation}
where $L_{\rm Edd}=1.4\times 10^{38}\left(\mbh/\sunm\right)$\ergs. Fig. 6 shows the Eddington ratio 
distributions. We find that there is a fraction of objects with super-Eddington accretion. This is caused by the 
low mass tail of the black holes arisen from the underestimation of the black hole mass. However we find the
Eddington ratios from \oiii method are similar in BLS1s and HBLR S2s at a level of $9.7\%$.
The mean Eddington ratios are $\langle\log\cale\rangle=-0.29\pm 0.09$ and $-0.49\pm 0.14$ 
for BLS1s and absorbed HBLR S1s, respectively, having sub-Eddington accretion. These results imply 
that the standard accretion disks are powering the central engines in BLS1s and HBLR S2s. 

\figurenum{6}
\centerline{\includegraphics[angle=-90,width=8.0cm]{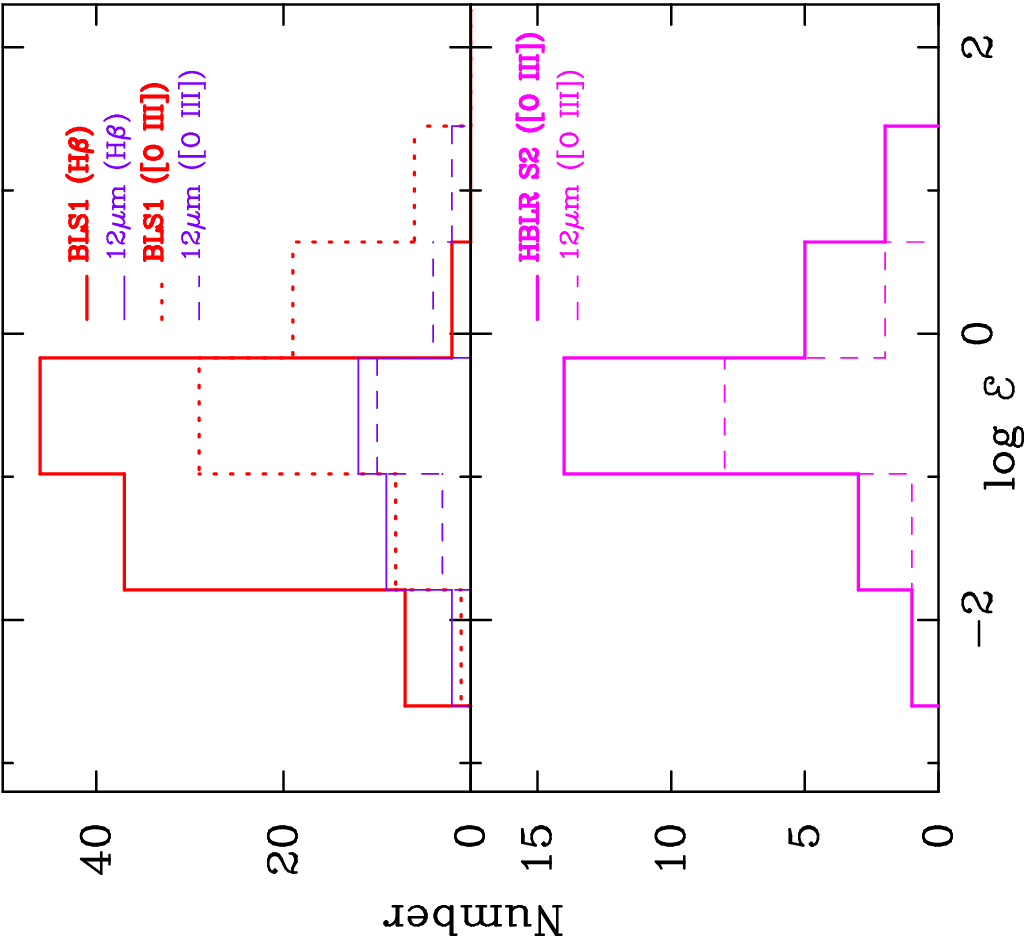}}
\figcaption{The distribution of the Eddington ratios of black holes in BLS1s and hidden broad line Seyfert 2
galaxies. We find that the Eddington ratios in both BLS1s and HBL S2s are similar to that in 12$\mu$m sample.}
\label{fig6}
\vglue 0.2cm

\subsubsection{Hard X-ray Spectrum}
Hard X-ray spectrum is a good indicator of the Eddington ratio in black hole accretion disks
(Wang et al. 2004; Shemmer et al. 2006). 
This allows us to estimate the Eddington ratios in HBLR S2s since there is less absorption in 2-10keV.
The physical reason for this correlation is: the seed photons can efficiently cool the hot electrons 
via Comptonization, steppening the hard X-ray spectrum (Wang et al. 2004). The larger the Eddington ratio,
the steeper hard X-ray spectrum. The Seyfert galaxy sample shows that the 2-10keV indies are 
$\langle \Gamma\rangle=1.51\pm 0.09$ with a dispersion 0.24 for HBLR 2s and $\langle 
\Gamma\rangle=1.76\pm 0.03$ with a dispersion 0.5 for Seyfert 1s. The large scatter of $\Gamma$ 
distribution in Seyfert 2 galaxies is caused by the fact that there is strong absorption in soft band, 
leading to the spectrum flatten. Though the mean value of $\Gamma$ is different from Seyfert 1 galaxies, 
the peak value is same. This implies that the Eddington ratio may be of same values in most BLS1s and 
HBLR S2s.  This reflects that the Eddington ratios have same range in Seyfert 1 and HBLR Seyfert 2 
galaxies, indicating the same accretion rates in BLS1s and HBLR S2s.

\subsection{The multiwavelength properties}
Tran (2003) gave a detailed comparison between BLS1s and HBLR S2s, showing that indeed they have similar 
properties from radio to X-rays. Since the present study is not an extension of Tran (2003), we do not compare 
the multiwavelength properties for the present sample. Future multiwavelength study of entire kinds of Seyfert 
galaxies is needed to get the global properties. We believe that they will share the same properties.
As a summary, the comparison of central engines is given in Table 7 for BLS1s and HBLR S2s. We would like to 
draw a conclusion that BLS1s and HBLR S2s have intrinsically same central engines: same black hole masses and 
accretion rates, which supports the orientation-based unification. 

\figurenum{7}
\centerline{\includegraphics[angle=-90,width=8.0cm]{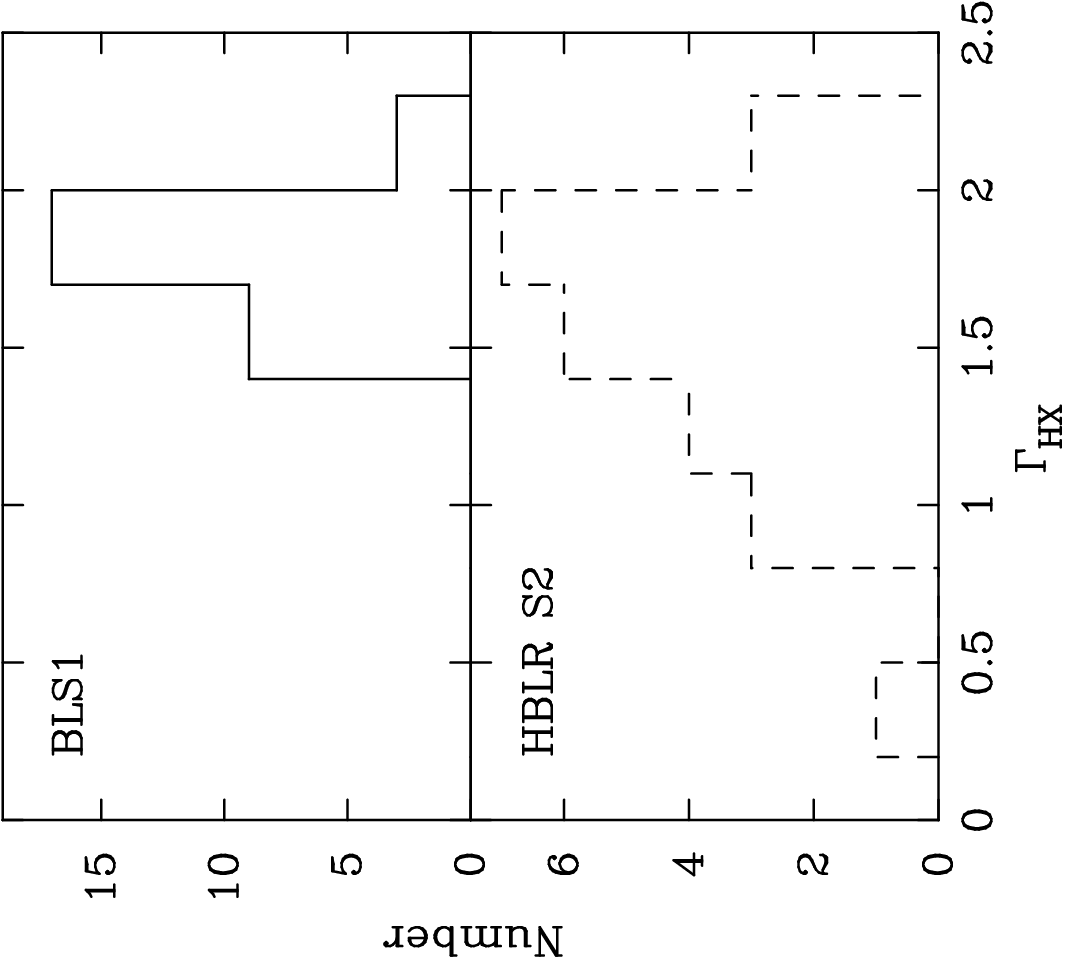}}
\figcaption{The distribution of the hard X-ray spectral index $\Gamma$. The peak values of the index 
$\Gamma$ are same in BLS1s and HBLR S2s. This indicates that the Eddington ratios could be similar. 
We note that the soft tail might be caused by the absorption of soft X-rays by torus.}
\label{fig7}
\vglue 0.2cm

\section{Unabsorbed Seyfert 2 Galaxies}
Figure 1 gives three kinds of unabsorbed Seyfert 2 galaxies, which are characterized by low or absent  
absorption in X-rays, and the black hole mass, respectively, but they have polarized broad lines (unabsorbed HBLR 
S2s), or no polarized broad lines (unabsorbed non-HBLR S2s). These Seyfert 2 galaxies, especially the unabsorbed 
non-HBLR S2s, are different from those absorbed HBLR S2s. Figure 8 displays the $N_{\rm H}$ distribution in Seyfert
2 galaxies. The definition of unabsorption is the hydrogen column density less than $10^{22}$cm$^{-2}$ (PB02), which
corresponds to a Thomson scattering depth of $10^{-2}$. 
Actually this is not a strict definition\footnote{Indeed the $N_{\rm H}-$distribution shows a significant fraction of 
low absorption less than $10^{22}$cm$^{-2}$, which is a reasonable value for distinguishing absorbed and unabsorbed
Seyfert 2 galaxies.}, but it indicates that the absorption is very weak.

The group of unabsorbed Seyfert 2 galaxies is not a minority. Mainieri et al. (2002) found that nearly 30\%
Seyfert 2 galaxies show $N_{\rm H}<10^{22}$cm$^{-2}$, also suggested by PB02. Five out of 28 ($\sim 17.8\%$) 
optically defined type II AGNs exhibited low levels of absorption in the deep {\em XMM}-Newton
observations of the Lockman Hole (Mateos et al. 2005). In the present sample, unabsorbed Seyfert 2 galaxies
occupy about $25.7\%$. Optically-defined Seyfert 2 galaxies show very weak absorption
along the line of observer's sight, which are exceptional in the orientation-based unification scheme.

\subsection{Black Hole Mass}
Table 5 gives the samples of the unabsorbed Seyfert 2 galaxies. We divided them into two types according to
whether they show polarized broad lines. However most of them have not been observed by spectropolarimetry. 
As we stated in the Section 2, we use Nicastro's criterion to classify them.
The first ones are those with brighter X-rays, such as the objects in G06. They are usually at the same level 
of Seyfert 1s in X-ray band (Gallo et al. 2006; Cappi et al. 2006).  The second are those with low luminosities 
of \oiii and X-rays, such as objects in PB02. Most of objects in PB02 are unabsorbed non-HBLR S2s.

We estimate the black hole mass via \oiii width if it is available, otherwise via
the dispersion velocity $\sigma$ or $\mbh-M_{\rm bulge}$ relation (H\"aring \& Rix 2004) if $\sigma$ is not
available from the literatures. When the torus is going to be exhausted,
the BLR is shrinking and the narrow line region is also shrinking according to the relation 
$R_{\rm NLR}=2.1L_{{\rm [O III]},42}^{0.52}$kpc, where $L_{{\rm [O III]},42}=L_{\rm [O III]}/10^{42}$\ergs 
(Bennert et al. 2002; Schmitt et al. 2003; Netzer et al. 
2004). For lowest \oiii luminosity $L_{\rm [O III]}\sim 10^{39}$\ergs, we have the NLR size is
$R_{\rm NLR}\approx 66$pc. Such a dimension of the NLR could be in the realm of the central black hole
(Laor 2003), the relation between $\sigma$ and \oiii could be broken so that $M_{\rm BH}-$FWHM(\oiii)
does not apply. We keep this in mind and still
use $\sigma=$\oiii/2.35 to estimate the black hole mass.  

Figure 9 shows the black hole mass distribution in unabsorbed non-HBLR S2s. We find it has a wide
distribution with a low mass tail ($\le 10^7\sunm$). This tail consists of three objects: 
NGC 1058, NGC 2685 and NGC 3486, especially NGC 1058 and NGC 3486, which have $\sigma\sim 60$\kms and bulge
magnitude $M_{\rm bulge}\sim -15$. Both of them have a very low 2-10 keV luminosity of $\le 10^{39.6}$\ergs
(Cappi et al. 2006) and low \oiii luminosity. They are "dwarf Seyfert nuclei" since their H$\alpha$ luminosity 
is fainter than $10^{40}$\ergs (Ho et al. 1997). Such a small black hole with low accretion rates might
represent a class of "seed" black hole in the galactic center. They could be unabsorbed non-HBLR S2sA predicted
in Table 1. Obviously it is worth studying further.
Except for the two galaxies, the mean value of the black hole mass $\langle \mbh\rangle=10^{7.8}\sunm$. The peak
mass is obviously larger than that in absorbed non-HBLR S2s, HBLR S2s, NLS1s, BLS1s as shown in Figure 3. This 
provides striking evidence for the black hole evolution from low to high mass in Seyfert galaxies. The growth of 
the black holes from NLS1s to unabsorbed Seyfert 2 galaxies is 
$\langle \mbh^{\rm u-NHBLRS2}\rangle/\langle \mbh^{\rm NLS1}\rangle\sim 100$. If the accreting mass 
is from the dusty torus, it requires that the torus should be of $10^8\sunm$. We lack the estimation
of the torus mass in Seyfert galaxies, but it could be in the range of PG quasars (Haas et al. 2002).
The torus mass is enough to fueling the black holes in Seyfert galaxies.

\figurenum{8}
\centerline{\includegraphics[angle=-90,width=8.0cm]{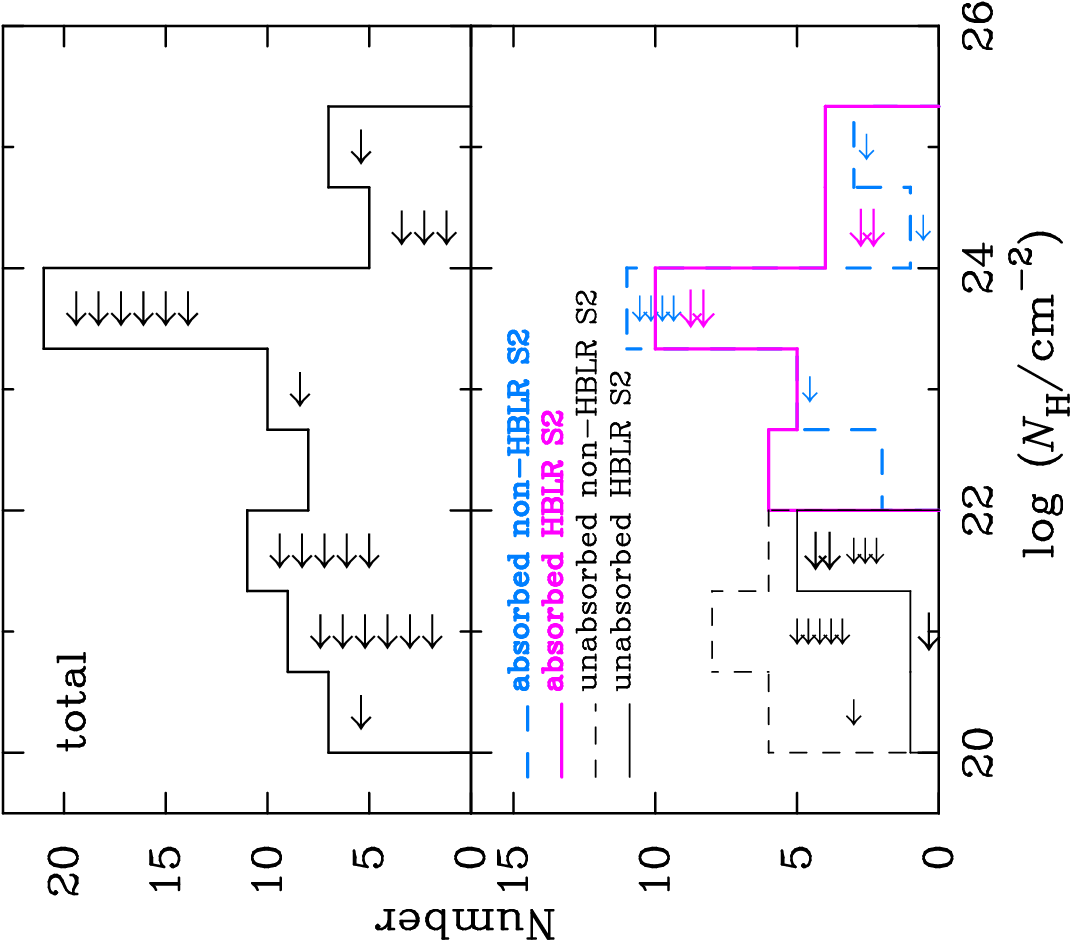}}
\figcaption{The $N_{\rm H}-$distrbution in {\em all} Seyfert 2, and absorbed/unabsorbed non-HBLR/HBLR S2s.
The definition of "unabsorbed" refers to $N_{\rm H}\le 10^{22}$cm$^{-2}$.}
\label{fig8}
\vglue 0.2cm

\figurenum{9}
\centerline{\includegraphics[angle=-90,width=8.0cm]{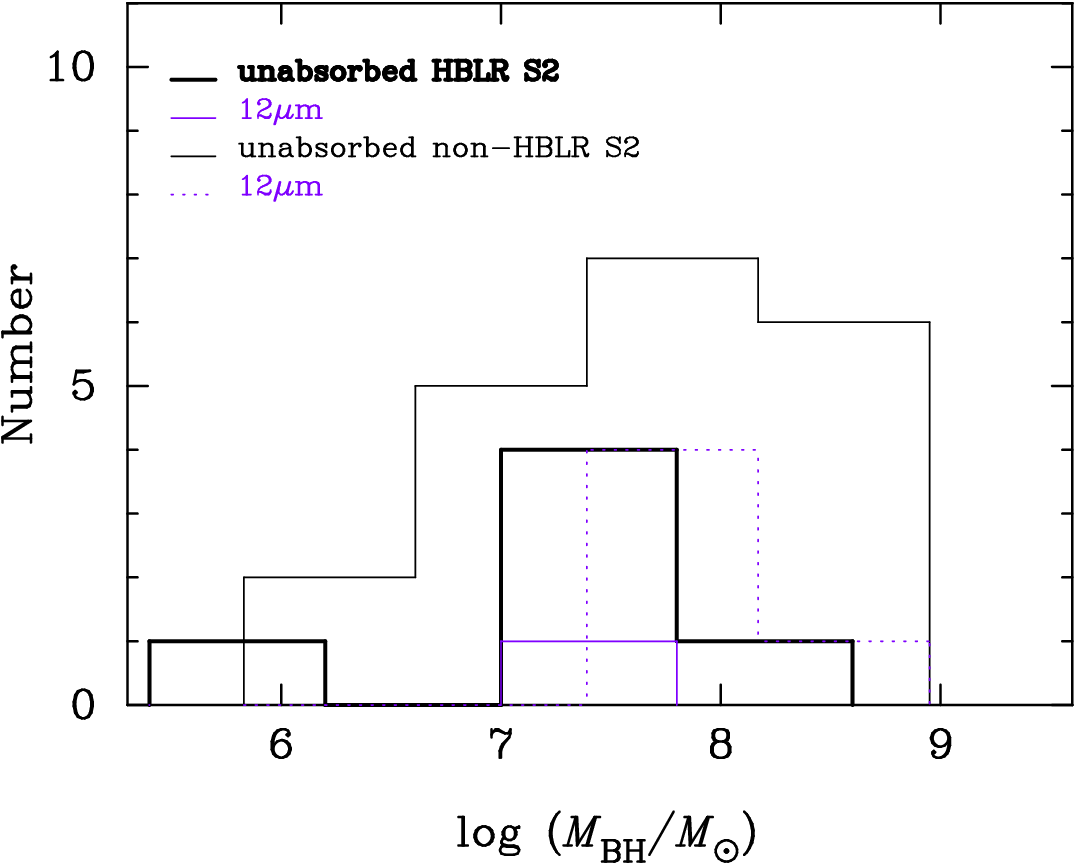}}
\figcaption{The distribution of the black hole masses in unabsorbed Seyfert 2
galaxies. }
\label{fig9}
\vglue 0.2cm

Figure 10 gives the distribution of bulge magnitudes in unabsorbed Seyfert 2 galaxies. The mean bulge magnitudes 
are $\langle M_{\rm Bulge}\rangle=-18.66$ and $\langle M_{\rm Bulge}\rangle=-18.73$ for the unabsorbed HBLR S2s
and non-HBLR S2s, respectively. Comparing Fig. 4 and 10, we find that unabsorbed Seyfert 2 galaxies have much 
broad distribution of bulge luminosities, with faint and bright bulge tails. The faint tail means that 
there is a group of unabsorbed Seyfert 2 galaxies which have less massive black holes. The low absorption in 
X-rays in these Seyfert galaxies means that the gas-to-dust ratio is low, indicating a dust-rich region. They may 
have undergone an intensive star formation process. We thus expect such a kind of objects is unabsorbed 
non-HBLR S2sB listed in Table 1. The bright 
tail indicates that the black holes are quite massive and have low Eddington ratios (see below section). They 
could be "true" Seyfert 2 galaxies with accretion rates below the critical value, reaching the final 
stage of Seyfert galaxies. We note that there is another potential explanation. More massive black holes with 
lower accretion rates is still visible, but the less massive black holes will drop out if they have lower accretion 
rates. So the luminosity-limit selection may ignore the less massive black holes with lower Eddington ratios.
However, the present sample with objects brighter than $L_{12\mu}>10^{42}$\ergs shows the only evolutionary way
from low to high mass.

\figurenum{10}
\centerline{\includegraphics[angle=-90,width=8.0cm]{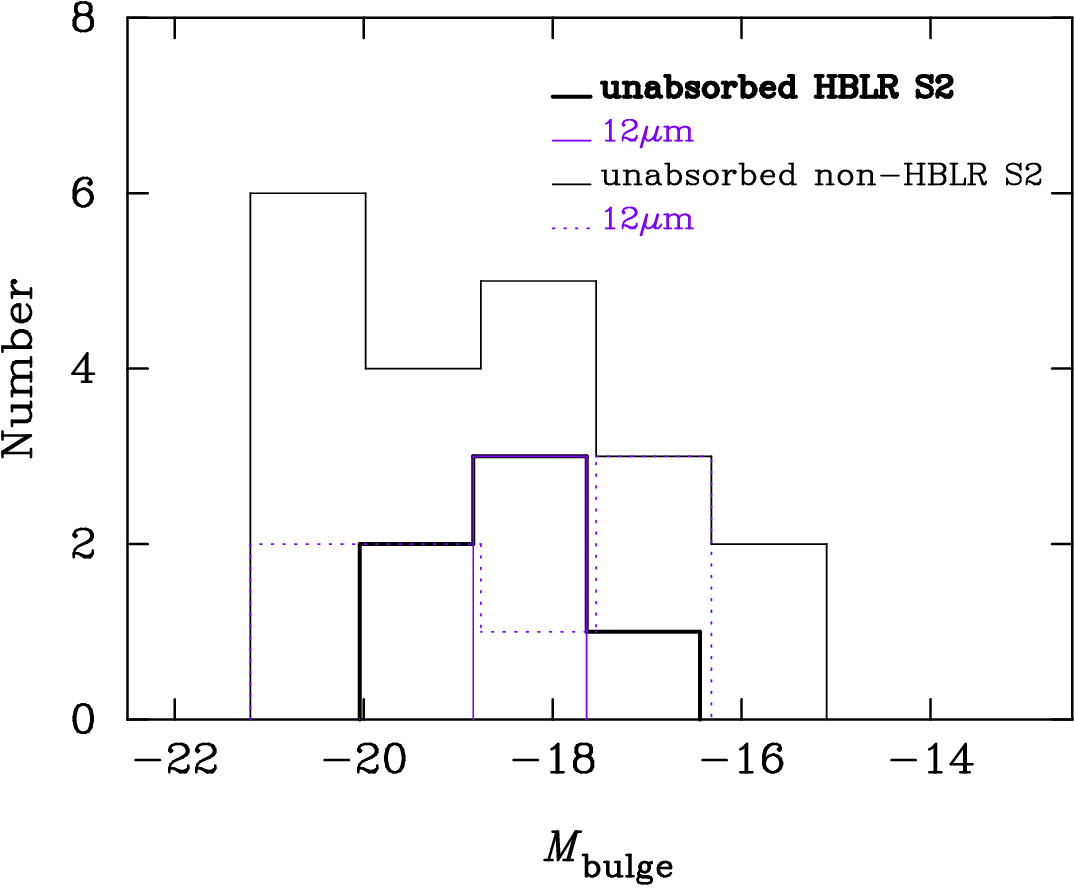}}
\figcaption{The distribution of the bulge magnitudes in unabsorbed Seyfert 2 galaxies. It shows that unabsorbed
non-HBLR S2s are significantly different from BLS1s and HBL S2s compared with Fig. 4. Comparing with NLS1s and 
absorbed non-HBLR S2s (Fig. 4 in ZW06), we find that the distributions of bulge luminosities in NLS1s, BLS1s, HBLR 
and non-HBLR S2s are very similar. This agrees with results of {\em Hubble} telescope that NLS1s hosts are similar 
to BLS1s (Crenshaw et al. 2003), but NLS1s have a bar more often than BLS1s.}
\label{fig10}
\vglue 0.2cm

\subsection{Accretion rate distribution}
Since the limitation of spectropolarimetric observations, only a few (only seven) of objects have been observed 
in the polarized light. We distinguished them via the Eddington ratios, but it is still interesting to display 
{\em how} the Eddington ratios distribute? We still use the relation $L_{\rm Bol}\approx 3500L_{5100}$ 
for unabsorbed Seyfert 2 galaxies. There are
some cases that the bolometric luminosity estimated from \oiii is very different from $L_{\rm 2-10keV}$.
When it happens, we use $L_{\rm Bol}=30L_{\rm 2-10keV}$ to estimate the Eddington ratio since we use
Nicastro's criteria (based on X-ray continuum). It is not clear the covering factor of the NLR in unabsorbed 
Seyfert 2s (regarding \oiii emission), most likely it is smaller than normal 
Seyferts. This gives an upper limit of the bolometric luminosities and the Eddington ratios.
The seven objects with spectropolarimetric observations are F01475-0740, NGC 2992, NGC 5995, NGC 3660,
NGC 4501, NGC5929 and NGC 7590. 
The first three objects are X-ray unabsorbed HBLR S2s, which have the mean
value of $L_{\rm Bol}/L_{\rm Edd}\approx 0.28$. The last four objects are X-ray unabsorbed non-HBLR S2s.
We find that they have Eddington ratios, $L_{\rm Bol}/L_{\rm Edd}\approx 0.03$, which is much lower than
that in the X-ray unabsorbed HBLR S2s. The Eddington ratio is shown in Figure 11. It is clear that most of the 
black holes have very sub-Eddington ratio ($<10^{-2}$) in unabsorbed non-HBLR S2s, peaking at $\cale\sim 10^{-3}$. 
This directly implies that the optically thin
advection-dominated accretion flows (ADAF) are powering these unabsorbed Seyfert 2 galaxies\footnote{We have to
address here that we draw such a conclusion, which is based on the four objects with spectropolarimetic 
observations.}.

The fraction of the unabsorbed Seyfert 2 galaxies to total Seyfert 2s
may be in a range of $10\%-30\%$, which could constitute of a special group of AGNs (PB02). In our sample 
the fraction of unabsorbed non-HBLR Seyfert 2s is ${\cal R}_F=7\%$ to total Seyfert galaxies. They could undergo a phase that
the clouds in the broad 
line region are fading away following decreases of accretion rates. Their fraction thus provide an estimation 
of the lifetime of unabsorbed Seyfert 2 galaxies though their nature is not sufficiently understood.

We have to stress that we classify the unabsorbed Seyfert 2 galaxies according to the Nicastro's criterion
since only small fraction of the unabsorbed Seyfert 2 galaxies have spectropolarization observation. Fig. 11 
shows the Eddington ratio distribution, but is artificial according to their
classification. On the other hand the estimation of the black hole mass
in these Seyfert 2 galaxies is based on different methods. This may lead to random scatters in the black hole
mass distribution, which is very hard to evaluate. Table 6 shows that the appearance frequencies of unabsorbed 
HBLR S2s are much lower than unabsorbed A non-HBLR S2s (6:20). This implies that unabsorbed HBLR S2s are 
short-lived since the intensive star formation exhausts gas. They are thus characterized by low 
gas-to-dust ratios. After a very short period of the unabsorbed
HBLR phase, they will finally evolve into "true" Seyfert 2 galaxies, which have too low 
Eddington ratios to have broad line region (Nicastro et al. 2003; Laor 2003).

\figurenum{11}
\centerline{\includegraphics[angle=-90,width=8.0cm]{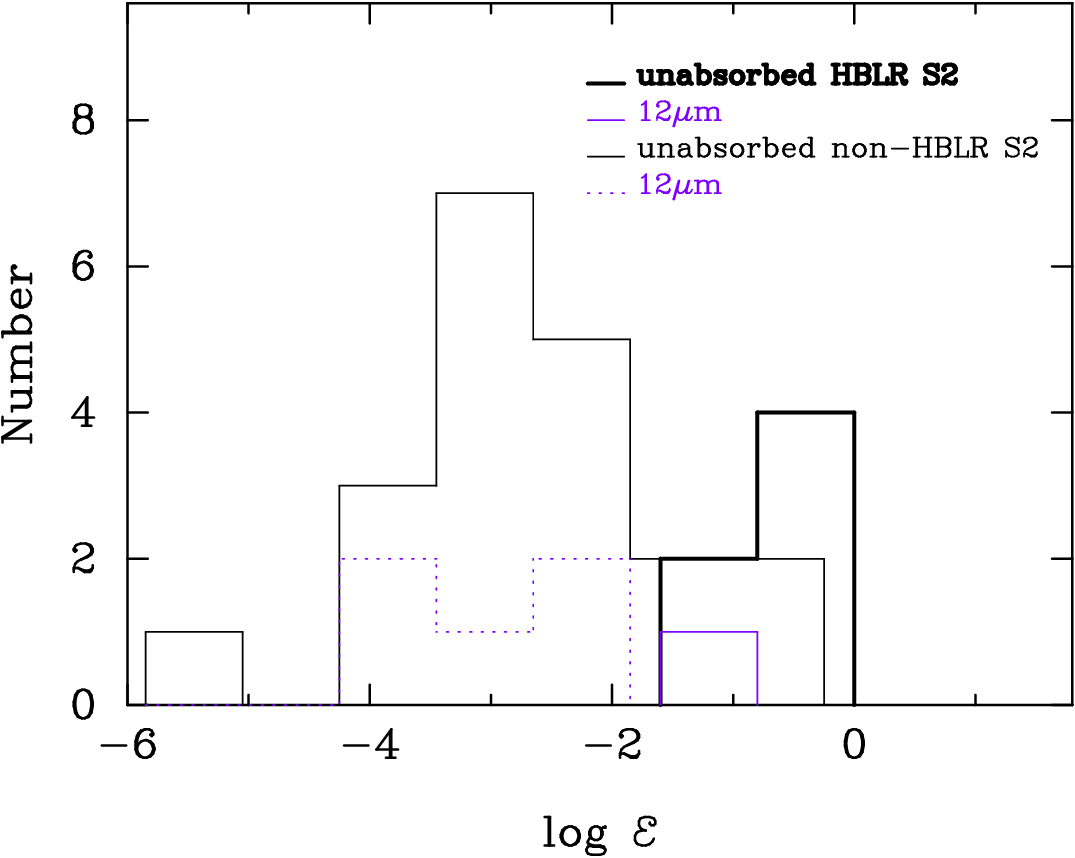}}
\figcaption{The distribution of the Eddington ratio in unabsorbed Seyfert 2 galaxies. }
\label{fig11}
\vglue 0.2cm

In this section, we show complex distributions of black hole masses in X-ray unabsorbed S2s and find that those 
objects with spectropolarimetric observations are in low accretion states. These are only preliminary results 
since the number of the sample is limited. Future work on this issue is needed.

\section{Nuclear Starburst in Seyfert 2s}
As argued by Maiolino et al. (2001a,b), Boller et al. (2003) and Gallo et al. (2006),
the gas-to-dust ratio inside the torus plays an important role in 
observational appearance of a Seyfert 2 galaxy. The composition of the torus is controlled by its star 
formation history. To test the roles of the gas-to-dust ratio, we should investigate the star formation 
rates and the hydrogen column density estimated from X-ray observations. 
It has been found that the star formation rates in Seyfert 1s are 
identical to that in Seyfert 2s galaxies from IRTF 3m telescope (Imanishi \& Wada 2004)
based on the 3.3$\mu$ features of polycyclic aromatic hydrocarbon (PAH). Here we focus on Seyfert 2 galaxies 
since the column density $N_{\rm H}$ roughly represents the gas column density. 

\figurenum{12}
\centerline{\includegraphics[angle=-90,width=8.0cm]{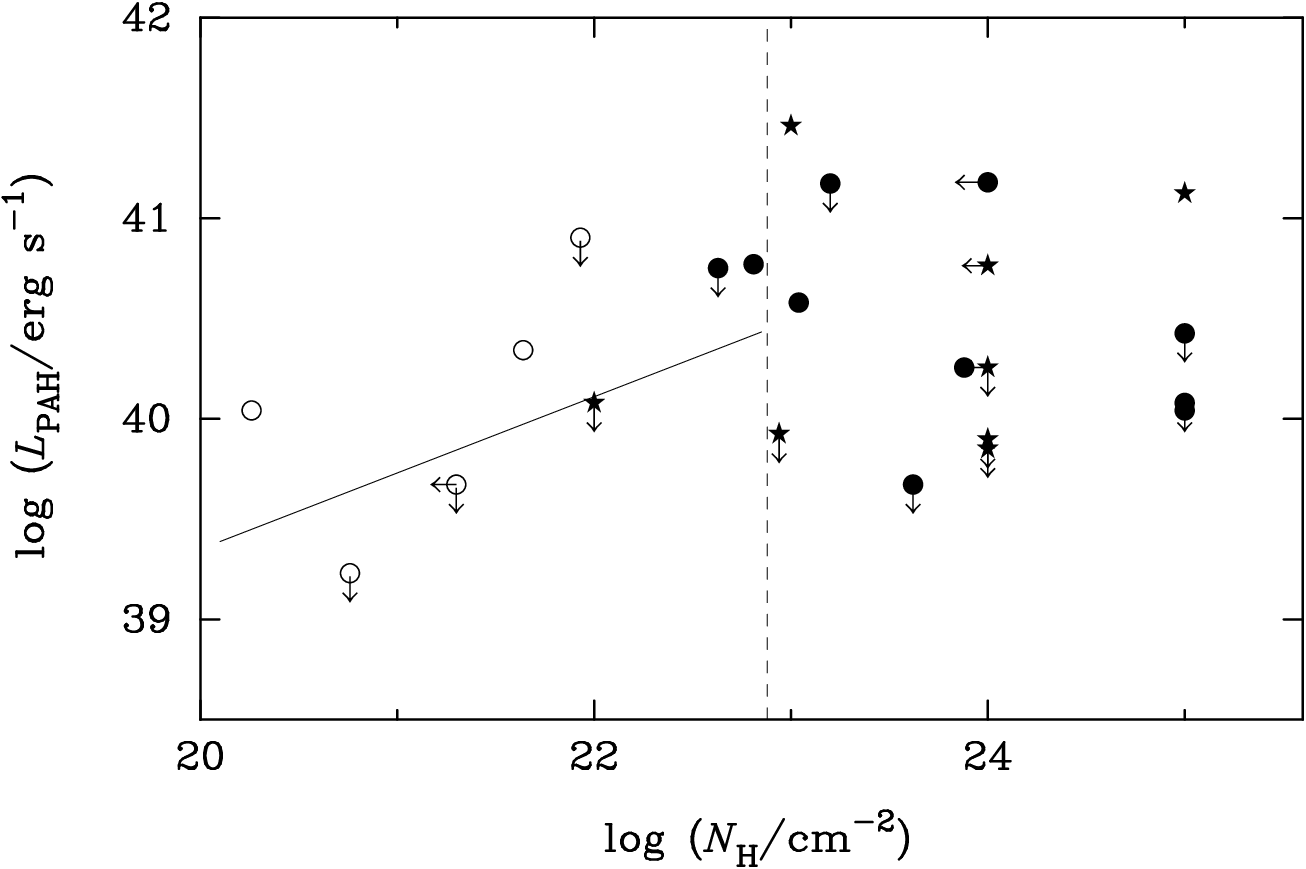}}
\figcaption{The plot of the 3.3$\mu$ PAH luminosity and hydrogen column density. The open circles are
unabsorbed Seyfert 2 galaxies, the solid circles the absorbed HBLR S2s in our sample and the
stars the absorbed non-HBLR S2s in ZW06 [the 3.3PAH luminosities are taken from Imanishi \& Wada (2004)].
The unabsorbed Seyfert 2s are those whose hydrogen column density is below $10^{22}$cm$^{-2}$.
}
\label{fig12}
\vglue 0.2cm

Imanishi (2002, 2003) showed that $3.3\mu$m PAH emission can represent the star formation rates in nuclear 
region. Figure 12 shows a plot of the $3.3\mu$m luminosity with the hydrogen column density obtained from X-ray 
observations. For those Seyfert 2 galaxies with low $N_{\rm H}$ ($<10^{22}$cm$^{-2}$), there is a clear 
trend\footnote{The ASURV analysis shows that there is a correlation between $L_{\rm PAH}$ and $N_{\rm H}$ 
as $\log L_{\rm PAH}=(0.38\pm 0.18)\log N_{\rm H}+(31.75\pm 3.88)$ with Spearman's $\rho=0.54$ and 
$p_{\rm null}=0.16$.} that the higher the $N_{\rm H}$ the higher the 3.3$\mu$ PAH luminosity. The PAH 
luminosity corresponds to IR luminosity $\sim 10^{42\sim 43}$\ergs (Imanishi 2002), implying a star formation 
rate $\sim 0.5\sunm {\rm yr^{-1}}$ around the central region of a few 100 pc (eq. 4 in Kennicutt 1998). The 
intensive star formation implies a dust-rich nuclear 
region. However the PAH luminosity tends to be a constant with $N_{\rm H}$ increases. On the one hand, we have to 
note that the PAH grains may be destroyed through sublimation by the strong radiation from the central engine 
(Voit 1992), showing a saturated PAH luminosity. We thus may underestimate the star formation rates. On the other 
hand, the suppression of star formation in the nuclear region could be caused by the activity of the black holes 
like in elliptical galaxies (Schawinski et al. 2006). The higher the column density $N_{\rm H}$, the higher 
number densities of the molecular clouds inside the torus and as well as the frequencies of collisions among the
clouds. The gas escaping from the collisions is losing their angular momentum and fueling to the black holes. 
The higher $N_{\rm H}$ will enhance the accretion rates of the black holes. It is thus expected to have more 
strong feedback to the nuclear star formation region in a higher $N_{\rm H}$ objects. We show an argument for the 
presence of strong feedback in Seyfert 2 galaxies in \S7.2. However it is hard for us to draw a profound conclusion 
based on the present sample on the feedback to star formation in the nuclear region. Future study on this subject 
will help understand physics of star formation in this region. We do not find any distinction between HBLR S2s and 
non-HBLR S2s, this confirms the conclusion of Cid Fernandes et al. (2004), who did not find the difference of star 
formation rates in absorbed HBLR S2s and non-HBLR S2s (see Fig. 12). This implies that only X-ray absorption is affected
by the star formation history. Additionally if the black holes are fueled from 
the torus (Krolik \& Begelman 1988), the star formation rates will decrease when the gas is exhausted, showing a low 
3.3$\mu$ luminosity.

To further study the physics of Fig. 12, we define a parameter 
\begin{equation}
q=\log\left(\frac{L_{\rm PAH}}{N_{\rm H}}\right).
\end{equation}
The parameter $q$ reflects the gas-to-dust ratio inside the torus. Since 
$L_{\rm PAH}\propto L_{\rm IR}\propto \Sigma_{\rm SFR}$, where $L_{\rm IR}$ is IR luminosity and
$\Sigma_{\rm SFR}$ is the surface density of the star formation rates, we have 
$q\propto \log \Sigma_{\rm SFR}/N_{\rm H}\propto \log \rho_{\rm dust}/\rho_{\rm gas}
\propto \log{\cal Q}$, where we define ${\cal Q}=\rho_{\rm gas}/\rho_{\rm dust}$ (see \S 7.2). 
The gas-to-dust ratio is then shown by the parameter $q$. Figure 13 shows the distribution of $q$ parameter 
in (23) Seyfert 2 galaxies with PAH observations. First, the parameter $q$ has a wide distribution, likely tends 
to have a potential bimodal distribution
though the current data does not allow us to have a robust conclusion. The $q-$distribution peaks at 
$16$ and $18$. Second, it is clear that the low$-N_{\rm H}$ objects tend to have larger $q$.
Since the boundary ($N_{\rm H}=10^{22}$cm$^{-2}$)
to distinct the absorbed and unabsorbed Seyfert 2 galaxies is artificial, it could be plausible that this
bimodal distribution corresponds to a physical separation of the absorbed and unabsorbed Seyfert 2 galaxies.
More data are needed to confirm the potential bimodal distribution.
This also provides evidence for the nature difference between the two kinds of Seyfert 2 galaxies.

\figurenum{13}
\centerline{\includegraphics[angle=-90,width=8.0cm]{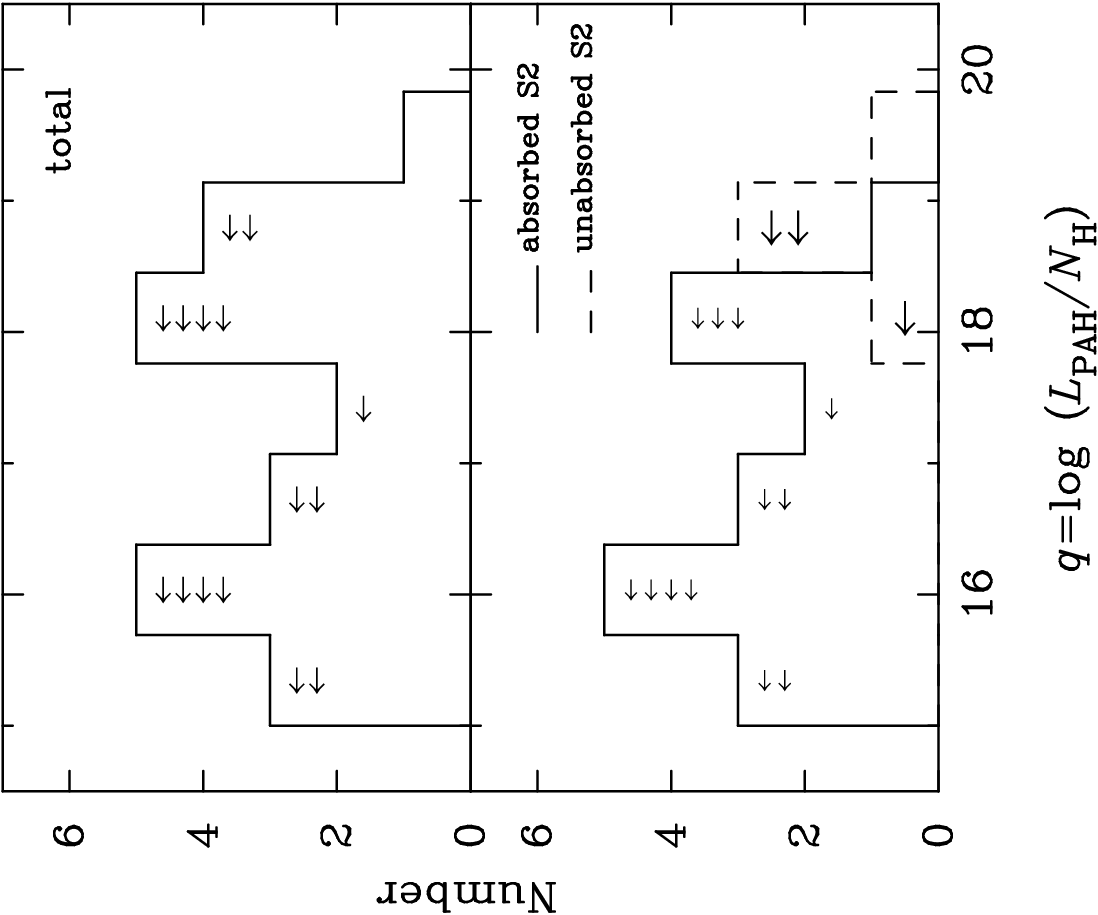}}
\figcaption{The distributions of the parameter $q$ in unabsorbed and 
absorbed Seyfert 2 galaxies. The solid line represents the absorbed Seyfert 2 galaxies and dotted line
unabsorbed Seyfert 2s. We find the parameter $q$ in absorbed Seyfert 2s are very different from in
unabsorbed Seyfert 2 galaxies. The results show that unabsorbed
Seyfert 2 galaxies have dust-rich nuclear region compared with absorbed
Seyfert 2 galaxies. The arrows in the plot only show the objects with the upper limits of PAH luminosities.
We do not include the $N_{\rm H}$ upper limits.
}
\label{fig13}
\vglue 0.2cm

\figurenum{14}
\begin{figure*}[t]
\centerline{\includegraphics[angle=-90,width=12.0cm]{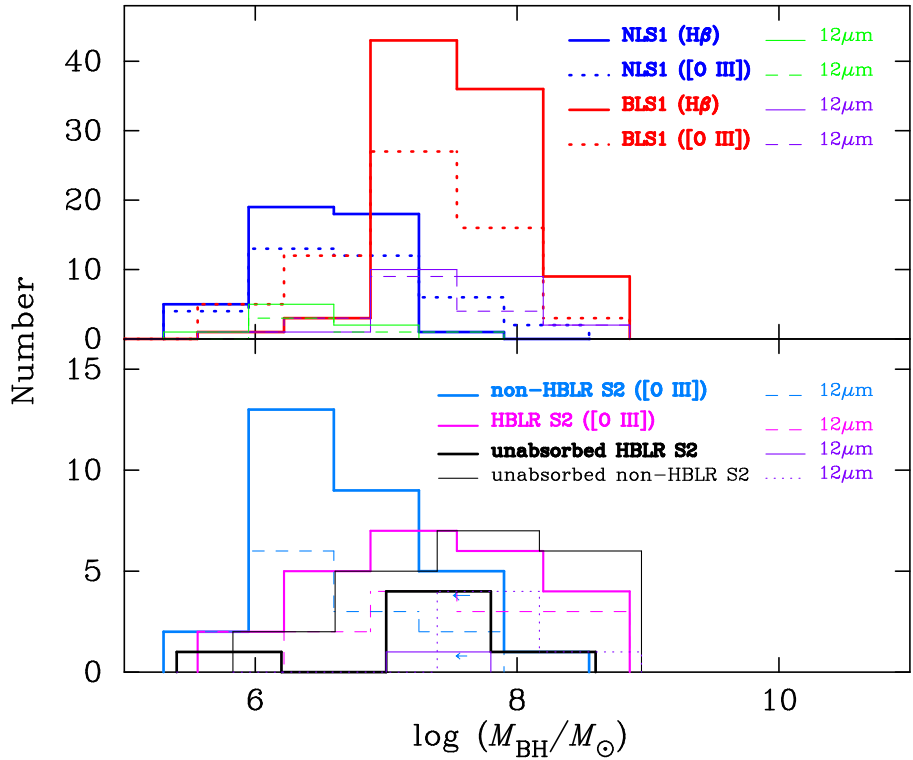}}
\figcaption{The plot of the distribution of black hole mass in non-HLBR S2s and NLS1s, 
HBLR S2s and BLS1s. It is clear that there are two peaks in the black mass distributions.} 
\label{fig14}
\end{figure*}
\vglue 0.2cm

There are several possibilities to explain the low absorption in optically-defined
Seyfert 2 galaxies (Boller et al. 2003). A faint broad line region is suggested
for an individual 1ES 1927+654, however we are not able to distinguish it from dust-rich model.
The present data show that the gas-dust
ratio ${\cal Q}$ is very different in absorbed and unabsorbed Seyfert 2s though the present sample is limited. 
The ratio ${\cal Q}$ is much less than that in absorbed Seyfert 2s, namely unabsorbed Seyfert 2
galaxies are dust-rich compared with absorbed Seyfert 2 galaxies. We draw such
conclusion with caution because of the sample is limited, however this 
confirms the dust-rich effects (Maiolino et al. 2001) and the hypothesis
that unabsorbed Seyfert 2s have intrinsically dust-rich nuclear region.

We note that the present sample is not large and some are upper limit sources of PAH flux. Future
test of the proposed scenario will be very important via a large sample of PAH observations by {\em Spitzer}.
We have to stress that the dust inside the torus, at least some of them, not only is produced by 
star formation, but also may be brought from outside of the  nuclear region.  It is worth studying 
the relation between the star formation history and hydrogen density 
inside the torus. This could be very helpful to uncover the relation 
between the fuel supplying and star formation in Seyfert evolution.

\figurenum{15}
\begin{figure*}[t]
\centerline{\includegraphics[angle=-90,width=11.5cm]{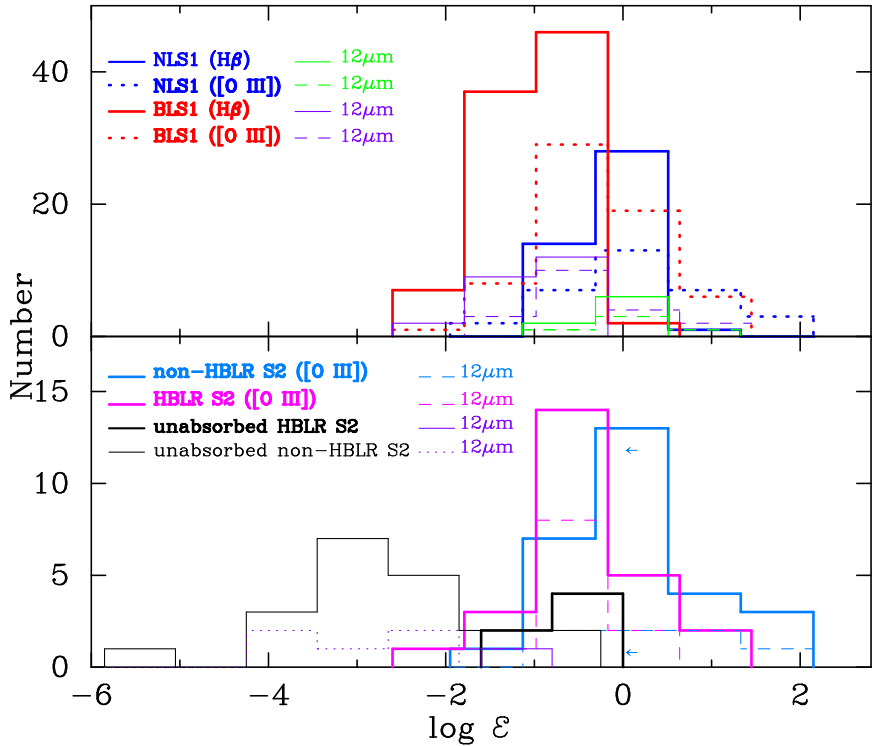}}
\figcaption{
The plot of the Eddington ratios in {\em all} kinds of Seyfert galaxies. We find that there is a 
sequence of the Eddington ratios among Seyfert galaxies with the same orientations.} 
\label{fig15}
\end{figure*}
\vglue 0.2cm

\section{The evolving central engines}
It is a popular opinion that galaxy interactions drive gas toward the galactic center, and then 
the gas, piled up enough, not only feeds the black hole but also triggers a starburst. Figure 1 
outlines a possible evolution sequence, but we do {\em not} mean that every 
BLS1s {\em must} pass through the stage of NLS1s. There is evidence 
for that stellar bars are more common in NLS1s than BLS1s (Crenshaw et al. 2003) and NLS1s 
indeed host less massive galaxies (Botte et al. 2004, but see a contrary conclusion from Botte 
et al. 2005). This means that {\em not} all BLS1s originate from NLS1s, however, NLS1s are growing
due to accretion and will inevitably undergo the phase of broad line Seyfert galaxies. Here we focus 
on the case of evolution from NLS1s to the final stage.

\subsection{Statistic properties of Seyfert galaxies}
We give a brief summary of Seyfert galaxies. Table 8 lists the black hole masses 
and Eddington ratios of  Seyfert galaxies. Fig. 14 shows the plot of the combinations of the black 
hole mass distributions in all kinds of Seyfert galaxies discussed in the present paper and ZW06. It is very 
interesting to find that the black hole masses in NLS1s are systematically smaller than in BLS1s as well in 
non-HBLR S2s and BLS1s, namely, 
$$\begin{array}{ll}
\langle M_{\rm BH}\rangle_{\left\{{\rm NLS1, a-n-HBLR S2}\right\}}\rightarrow \\
~~~~~~~~~~ \langle M_{\rm BH}\rangle_{\left\{{\rm BLS1, HBLR S2}\right\}}\rightarrow\\
~~~~~~~~~~~~~~~~~~~~\langle M_{\rm BH}\rangle_{\left\{{\rm u-n-HBLR S2}\right\}}, 
\end{array}
$$
where a-n-HBLR S2 is absorbed non-HBLR S2s and u-n-HBLR S2 is unabsorbed non-HBLR S2. 
Figure 15 shows that the Eddington ratio distributions of the entire sample. We find a 
queue\footnote{This is only a preliminary result. We would like
to stress again that the low Eddington ratios deduced in X-ray unabsorbed non-HBLR S2s results
from their definition. However the four X-ray unabsorbed non-HBLR S2s indeed show they have very low
Eddington ratios. } 
$$
\begin{array}{ll}
\langle {\cal E}\rangle_{\left\{{\rm NLS1,a-n-HBLR S2}\right\}}\rightarrow  \\
~~~~~~~~~~ \langle {\cal E}\rangle_{\left\{{\rm BLS1, HBLR S2}\right\}}\rightarrow \\
~~~~~~~~~~~~~~~~~~~~\langle {\cal E}\rangle_{\left\{{\rm u-n-HBLR S2}\right\}},
\end{array}
$$
showing a decreasing trend.
These are  evidence for evolutionary connection in Figure 1. With the evolution of the black 
holes, we are able to tackle: 1) the changing opening angle of the torus; 2) the lifetimes determining
the relative numbers of the Seyfert galaxies. This will give global description of Seyfert evolution.

The present sample has good completeness and homogeneity like 12$\mu$m sample as shown in Table 6,
which gives the relative numbers of the five kinds of Seyfert galaxies. As we argued in previous sections,
the diversity of Seyfert galaxies makes the relative numbers inconsistent among different band surveys, but
only within a few percent, at maximum 10\%.
Except for NLS1s, the deep survey of the Lockman Hole with the {\em XMM}-Newton
show nearly $30\%$ of type II Seyfert galaxies have low absorption with $N_{\rm H}<10^{22}$cm$^{-2}$
(Mainieri et al. 2002), however the 12$\mu$m, Tran (2003) and the present samples
give about $\sim 10\%$. We add each kind
objects from the three samples and then get the mean fraction by dividing the total numbers of the three 
samples so as to avoid the incompleteness and inhomogeneities as we can. 
We will only use the mean fractions in the following sections.

\figurenum{16}
\begin{figure*}[t]
\centerline{\includegraphics[angle=-90,width=15.5cm]{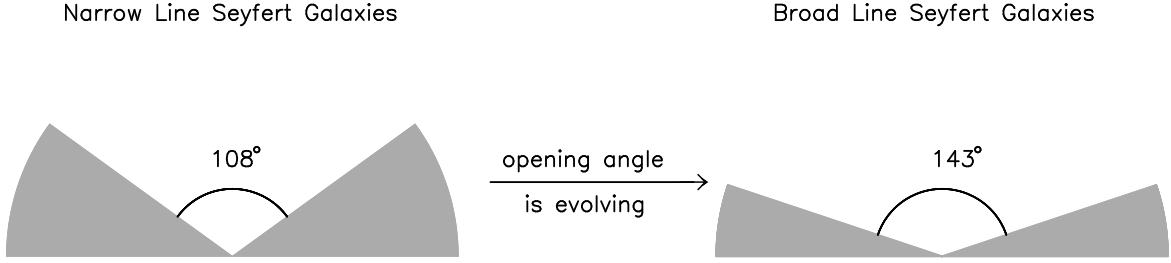}}
\figcaption{
A carton shows the evolving opening angle along with growth of black holes. This carton only shows upper
section of the torus in Seyfert galaxies. We find that the opening angle of the torus is increasing with
growth of the black hole from the present sample. The torus here is referred to the compact region, rather 
than the extended region, where the star formation dominated (Imanish \& Wada 2004). } 
\label{fig16}
\end{figure*}
\vglue 0.2cm

\subsection{Unified model with evolution} 
The relative numbers of Seyfert 2 galaxies could provide the opening angle of the dusty torus (Osterbrock 
\& Shaw 1988; Salzer 1989; Huchra \& Burg 1992; Wilson \& Tsvetanov 1994; Maiolino \& Rieke 1995; Schmitt 
\& Kinney 2000; Tovmassian 2001). These previous studies on the ratio do {\em not} distinguish
the differences between HBLR S2s and non-HBLR S2s, whose central engines are intrinsically different. 
The opening angles are estimated incorrectly in the previous studies. The present sample allows 
us to calculate the opening angles for two pairs of (BLS1s and HBLR S2s) and (NLS1s and 
non-HBLR S2s), which are different only purely for orientations. Assuming a constant geometry
of the dusty tori for each subclasses, we have the opening angle
\begin{equation}
\cos\left(\frac{\Theta_N}{2}\right)=\frac{N_{\rm NHBRS2}}{N_{\rm NHBRS2}+N_{\rm NLS1}},
\end{equation}
and
\begin{equation}
\cos\left(\frac{\Theta_B}{2}\right)=\frac{N_{\rm HBRS2}}{N_{\rm HBRS2}+N_{\rm BLS1}},
\end{equation}
for (NLS1s, non-HBLR S2s) and (BLS1s, HBLR S2s), respectively. The subscripts $N$ and $B$ refer to the 
narrow/broad line Seyfert galaxies.  If we 
establish an homogeneous sample consisting of four kinds of Seyfert galaxies, we then have opening angles 
$\Theta_N$ and $\Theta_B$ of the dusty tori, respectively, and get some indications of the torus evolution 
along with the growth of the black holes. We take the data 
from Table 6. NLS1s occupy $14\%$ of the total Seyfert galaxies and absorbed non-HBLR S2s
$20\%$, we have $\cos\left(\Theta_N/2\right)=0.59\pm 0.10$, where the uncertainty is obtained by assuming $\sim 5\%$ 
uncertainties of each sub-classes. For broad line Seyfert galaxies, the fraction of HBLS2s is of
$18\%$ and BLS1s $38\%$ from Table 6, we have $\cos\left(\Theta_B/2\right)=0.32\pm 0.09$. Finally we have the opening 
angles
\begin{equation}
\Theta_N\approx {107.6^{\circ}}\pm 14.2^{\circ};~~~~~\Theta_B\approx {142.7^{\circ}}\pm 8.0^{\circ},
\end{equation}
for NLS1 and BLS1s, respectively. $\Theta_B$ is obviously
different from the previous results (e.g. Osterbrock \& Shaw 1988; Tovmassian 2001) since these authors
did not distinguish HBLR S2s and non-HBLR S2s. We find $\Theta_N<\Theta_B$ at a level $>3\sigma$. We address here 
the significant differences between the narrow and broad line Seyfert galaxies\footnote{Since the appearance of NLS1s 
and non-HBLR S2s is a little uncertain from Table 6, their opening angles should be treated with a little more caution
than that in broad line Seyfert galaxies. This should also be noted for ${\cal R}_N$ in \S6.3}. This result
provides one piece of new evidence for the evolving torus: the material from torus 
is supplying to the central black hole and then getting the covering 
factor decreased as originally suggested by Krolik \& Begelman (1988). Considering
accretion rates in NLS1s and absorbed non-HBLR S2s are systematically higher than 
that in BLS1s and HBLR S2s, we could draw a conclusion that NLS1s and the non-HBLR
S2s are young AGNs. This agrees with the conclusions from radio emission (Komossa et al. 2006).
The opening angles of torus in NLS1s and absorbed non-HBLR S2s are decreasing with
the growth of the black hole due to accretion from the dusty torus.

\subsection{Accretion: growth of black holes}
However only supplement of the receding opening angle of the torus to the standard unification scheme is 
not enough. Accretion onto the black holes significantly increases 
their masses (Kawaguchi et al. 2004). The evolutionary unification scheme should be studied so that 
we can reveal the history of black hole growth due to accretion.
This may allow us to set up the evolutionary sequences of the activities.
Mathur (2001) found that the mean black hole to bulge mass ratio of NLS1s
is significantly smaller than that for normal Seyfert galaxies. The ratio 
of black hole mass to bulge velocity dispersion is also significantly 
smaller for NLS1s. This supports the conjecture that the NLS1s are young AGNs.
The black hole would grow with higher accretion rates in the early stage and 
eventually the accretion rates and the growth would slow down.
BLS1s are probably evolutionary consequence of  NLS1s as they have more massive 
black holes and similarly, non-HBLR S2 may grow up to HBLS2s with the assumption 
that the orientation of the torus dose not change during evolution.

\figurenum{17}
\begin{figure*}[t]
\centerline{\includegraphics[angle=-90,width=15.5cm]{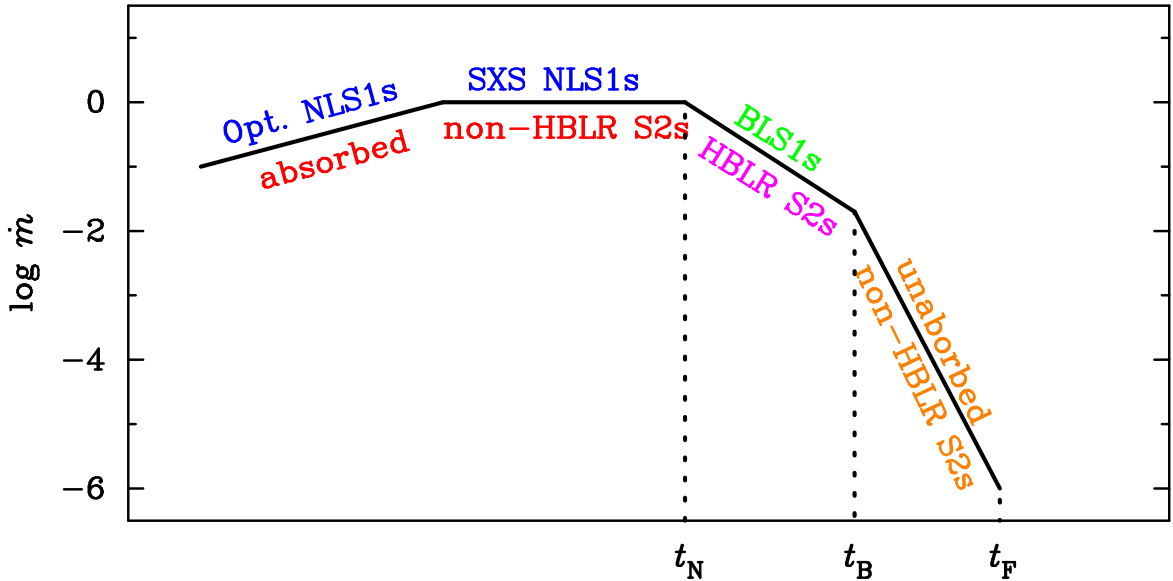}}
\figcaption{
The plot of the accretion rate evolution sequence in Seyfert galaxies. Opt. NLS1s: 
optically-selected NLS1s; SXS NLS1s: soft X-ray selected NLS1s. The three critical 
time intervals $t_N$, $t_B$ and $t_F$ are estimated from the relative number of the objects 
(see text for details). The plot is not scaled in time.} 
\label{fig17}
\end{figure*}
\vglue 0.2cm

We define the relative numbers of 
\begin{equation}
{\cal R}_N=\frac{N_{\rm NLS1}}{N_{\rm BLS1}};~~~~~
{\cal R}_B=\frac{N_{\rm NHBRS2}}{N_{\rm HBLRS2}},
\end{equation}
which will give the lifetime of NLS1s/non-HBLR S2s. If the evolutionary chains are parallel: 
non-HBLS2s$\rightarrow$HBLRS2s and NLS1s$\rightarrow$BLRS1s,
there then is a single relation  ${\cal R}_N={\cal R}_B$.  With data from Table 6, we have 
${\cal R}_N=0.37$ and ${\cal R}_B=1.0$. This suggests that the opening angle will {\em not} keep 
a constant during the growth evolution, confirming evolution of the opening angles. Equation (11) shows 
that the opening angle is evolving with growth of black holes, and there should appear across evolution: 
non-HBLS2s$\rightarrow$BLS1s, but have not NLS1s$\rightarrow$HBLRS2s. The present ${\cal R}_N < {\cal R}_B$ 
also indicates such a cross evolution. 
Since the torus geometry is indeed complicate, such a hybrid evolution could not be avoided. 
Considering such a complicate situation, we define a new parameter
\begin{equation}
{\cal R}_0=\frac{N_{\rm NLS1}+N_{\rm NHBRS2}}{N_{\rm BLS1}+N_{\rm HBLRS2}},
\end{equation}
which allows us to determine the relative number of the two groups of narrow and broad line Seyfert
galaxies and describes the evolutionary sequence as the first approximation. From Table 6, we have
${\cal R}_0=0.58$, implying that about $37\%$ Seyfert galaxies stay the narrow line stage.

Assuming an exponential growth of the black holes, we have their masses
at time $t$
\begin{equation}
\mbh(t)=\mbh^0\exp\left(\frac{\dot{m} t}{\eta t_{\rm Salp}}\right),
\end{equation}
where $\mbh^0$ is its initial mass, the Salpeter time $t_{\rm Salp}=0.45$Gyr,
$\dot{m}=\dot{M}/\dot{M}_{\rm Edd}$ and $\dot{M}_{\rm Edd}=L_{\rm Edd}/c^2$,
$L_{\rm Edd}$ is the Eddington limit and $\eta$ is the radiative efficiency.
If NLS1s/non-HBLR S2s are evolving into BLS1/HBLR S2s, the lifetimes of 
NLS1/non-HBLR S2s are then given by
\begin{equation}
t_N=t_{\rm Salp}\ln\left[\frac{\langle M_{\rm BH}^{\rm BLS1}\rangle}
               {\langle M_{\rm BH}^{\rm NLS1}\rangle}\right]^{\eta/\langle\dot{m}\rangle},
\end{equation}
where $\langle M_{\rm BH}^{\rm BLS1}\rangle$ and $\langle M_{\rm BH}^{\rm NLS1}\rangle$ are the mean mass of 
the black holes in BLS1s and NLS1s, respectively. From ZW06, we have 
$\langle M_{\rm BH}^{\rm NLS1}\rangle=10^{6.53}\sunm$ and $\langle\dot{m}\rangle=10^{0.19}$, 
then we have $t_N=6.7\times 10^7$yr with $\langle M_{\rm BH}^{\rm BLS1}\rangle=10^{7.53}\sunm$ from Table 8.
Considering the relative number of BLS1s/HBLRS2s to NLS1s/non-HBLR S2s, we then have the lifetime of BLS1s/HBLR 
S2s from
\begin{equation}
t_B=\frac{t_N}{{\cal R}_0}.
\end{equation}
The present sample provides $t_B=1.2\times 10^8$yr for the broad line Seyfert 1 galaxies.
Finally the lifetime of the unabsorbed non-HBLR Seyfert 2 galaxies can be determined by their
relative numbers ${\cal R}_F$ to BLS1s/HBLR S2s, we have
\begin{equation}
t_F={\cal R}_Ft_B=\frac{{\cal R}_F}{{\cal R}_0}t_N,
\end{equation}
we have $t_F=8.0\times 10^6$yr. Such a short timescale implies that unabsorbed non-HBLR S2s are rare objects 
in the universe\footnote{It should be realized that the lifetime $t_F$ refers to those unabsorbed non-HBLR S2s
brighter than $L_{12\mu}>10^{42}$\ergs. The very low luminosity AGNs are poorly unknown compared with Hao et al. 
(2005) sample (see below).}. After this time, the Seyfert 2 galaxies evolve into low luminous AGNs (of course, 
most of the 
unabsorbed Seyfert 2 galaxies are low luminous AGNs). This is consistent with the result from X-ray background 
constraints that AGNs have very short ADAF phase, appearing as low luminosity AGNs (Cao 2005). Indeed 
there is evidence for dust-free in low luminosity AGNs (LLAGNs), powered by ADAF or RIAF (radiative inefficient
accretion flow), which have quite "clean" environment 
in the nuclear region. The unabsorbed non-HBLR S2s stay there as long as the galaxies can not be triggered through 
galaxy interaction or tidal disruption by the black holes.

Regarding the nature of the unabsorbed S2sB (see Table 1), 
it is hard to find a robust reason why they have much lower \oiii luminosities compared with 
other Seyfert galaxies if they are caused by the obscuration 
of the dusty torus. The diminishing torus becomes so thin that it becomes transparent to the BLR emission 
lines. However the clouds in the previous BLR are disappearing according to $R_{\rm BLR}-L$ relation (Kaspi et al. 
2005), but the dust grains maybe extinct the broad emission lines since the evaporation radius is also 
shrinking (Laor \& Draine 1993). At the same time the narrow line region is contracting according to $R_{\rm NLR}-L$ 
relation (Netzer et al. 2004), the clouds surviving still in NLR will emit significantly broader \oiii than BLS1s. 
However the comparison is not straightforward since the outflow in the BLS1s and HBLR S2s strongly influences the 
width of the \oiii profile whereas the weak/absent outflow from the unabsorbed non-HBLR S2sB less affects 
\oiii width. We prefer a scenario that the unabsorbed non-HBLR S2sB are dying Seyfert galaxies and will 
finally evolve into LLAGNs.

Finally we have to point out that the relative number ${\cal R}_F$ of the X-ray unabsorbed non-HBLR S2s 
could be flexible because of the uncertain completeness of the very low luminosity AGNs (fainter than 
$L_{12\mu}=10^{42}$\ergs) compared with
Hao et al.'s sample. Unfortunately, the sample is totally unknown only the \oiii luminosity function.
The very low luminosity could be caused either by  low accretion rates or by less massive low black holes.
Future work on this sample should be done for its multiwave continuum and emission line properties and will
enhance the global evolution of Seyfert galaxies.

\section{Discussions} 
\subsection{Seyfert evolution}
We argue there are four parameters controlling the appearance of AGNs in the unification 
scheme of Seyfert galaxies: 1) orientation and opening angle of the torus; 2) black hole mass; 3) accretion
rates; 4) gas-to-dust ratio in the torus. According to the present data, we are suggesting a possible 
evolutionary sequence as shown in Fig 17. 

As we have shown that the black hole masses in NLS1s are the minimum, the evolution of Seyfert 
galaxies should start from them. Additionally the Eddington ratios in optically-selected NLS1s are
less than soft X-ray selected NLS1s. We think the optically-selected NLS1s are the starting point
and then is soft X-ray selected NLS1s according to their Eddington ratios.
The black hole masses are smaller and as well as the
accretion rates since it is just triggered. The Williams' sample is not
included in the present sample since most of their redshifts are larger
than 0.05. X-ray observations show they have a flat soft X-ray spectrum (Williams 
et al. 2004), their masses are at the same order of soft X-ray selected NLS1s based on the empirical 
reverberation relation and the Eddington ratios are of $\sim 0.5$ (Williams et al. 2004). Additionally, 
radio-loud narrow line Seyfert 1 galaxies are rather rare (Greene \& Ho 2005; Komossa et al. 2006).
One possible reason for this is the presence of slowly spinning black holes in NLS1s since they
could be just in the fast growing phase (Komossa et al. 2006).

The accretion significantly increases their black hole masses then. The total mass of the dusty torus 
is poorly known, but should be larger enough for fueling the black holes.  Since the viral timescale of the
clouds in the BLR is quite short compared with that of the black hole growth, the width of the Balmer 
lines will follow the potential of the growing black holes. The Balmer emission lines will become much broader 
along with the growth of the black holes. The strong radiation field will bloat evaporate the dust to form
the inner edge of the torus,
the electron scattering mirror will formed likely in this way (extensive discussions on the mirror is beyond
the scope of the present paper). The obscured broad lines will be scattered off to observers in the polarized
light as seen in the classical Seyfert 2 galaxies. 

With the growth of the central black holes, the dusty torus will be finally exhausted (Krolik \& Begelman 1988). 
The galaxies will finally evolve into the realm of unabsorbed non-HBLR S2sB, which show less 
absorptions in soft X-ray band. This is evidenced by the fact that their black hole masses are the maximum and 
the Eddington ratios are the smallest among the Seyfert galaxies. It is expected 
that advection-dominated accretion flows are powering the nuclei, in which the radio-loudness may be in 
radio-loud domains as suggested by Ho (2002). It is then expected that they could 
be radio-loud according to the radio-loudness. Future studies on their radio properties will be carried out. 

\subsection{On gas-to-dust ratio in Seyfert 2s}
Figure 13 implies that the gas-to-dust ratios are different between the absorbed and unabsorbed Seyfert 2 galaxies.
According to the simple model of dust grain, Gorenstein (1975) derived a relation [his eq. (4)]
\begin{equation}
{\cal Q}=\frac{\rho_{\rm gas}}{\rho_{\rm dust}}=3.6\times 10^3\left(\frac{N_{22}}{A_v}\right),
\end{equation}
where $N_{22}=N/10^{22}$cm$^{-2}$, $A_v$ is the extinction coefficient, $\rho_{\rm gas}$ and $\rho_{\rm dust}$ 
are the densities of the gas and dust grains. We call the ratio ${\cal Q}$ the gas-to-dust ratio of the torus. 
For typical optically-defined Seyfert 2 galaxies, $A_{\nu}=2-5$, we have 
${\cal Q}_c=(0.9-1.0)\times 10^3N_{22}$ for $A_{\nu}=3-4$. For ${\cal Q}>{\cal Q}_c$,
an optically-defined Seyfert 2 appears as an absorbed Seyfert 2, otherwise as an unabsorbed Seyfert 2 galaxy.
Applying Schmidt's law $\Sigma_{\rm SFR}=2.5\times 10^{-4}\Sigma_{\rm gas}^{1.4}$ (Kennicutt 1998), we can get the
dust mass $M_{\rm dust}=\pi R^2\Sigma_{\rm SFR} t\xi$, where $t$ is the duration time of star burst, $\xi$
is the percentage of the star converted into dust. The gas surface density can be obtained from
$\Sigma_{\rm gas}=75N_{22}~\sunm/{\rm pc^2}$, we then have
\begin{equation}
{\cal Q}=\left\{\begin{array}{ll}
7.1\times 10^2~N_{22}^{-0.4}t_8^{-1}\xi_{-2}^{-1}& {\rm (unabsorbed~ S2)},\\
&\\
1.1\times 10^2~N_{24}^{-0.4}t_8^{-1}\xi_{-2}^{-1}& {\rm (absorbed~ S2)},
\end{array}\right.
\end{equation}
where $\Sigma_{\rm SFR}$ and $\Sigma_{\rm gas}$ are the surface densities of the star formation rates and gas in 
unit of $\sunm$yr$^{-1}$kpc$^{-2}$ and $\sunm$pc$^{-2}$, respectively, 
$t_8=t/10^8$yr, $\xi_{-2}=\xi/10^{-2}$ (Sugerman et al. 2006). We find that both absorbed and unabsorbed Seyfert 2s have
${\cal Q}<{\cal Q}_c$ from equation (19), namely the nuclear region is gas poor.
This result agrees with X-ray observations of unabsorbed Seyfert 2s, however it directly conflicts with the 
strong absorption in absorbed Seyfert 2s. The only way to solve this problem
is that the Schmidt's law is broken in absorbed Seyfert 2 galaxies, in which 
the law predicts too much dust than observations. Why is the law broken in the nuclear region?
We think that the only cause is the activity in the nuclear region of Seyfert galaxies. The strong feedback
from the activity of the black holes efficiently suppresses the star formation in the nuclear region so as to
decrease the amount of dust. This simple argument shows the presence of feedback 
if the gas-to-dust ratio determines the X-ray appearance of an optically-defined Seyfert 2 galaxies.

There are plenty of feedback mechanisms operating in the nuclear region of Seyfert galaxies. The mainly efficient
feedback processes 
are Compton heating (Wang et al. 2005) and dissipation of the outflow's kinetic energy (Begelman 2004). The very 
strong nuclear outflows are indicated by the complicate profile of C {\sc iv} emission lines in NLS1s (Leighly 2000) 
and BLS1s (Crenshaw et al. 2003). The kinetic luminosity is of $10^{40-42}$\ergs inferred from X-ray warm absorber
in Seyfert galaxies (Blustin et al. 2005), which is enough to have strong feedback to their galaxies, at least
the nuclear region within the 
lifetime of NLS1s and BLS1s. The growth of the black holes will be indeed very important. The outflow from the 
center will finally strongly influence their hosts and keep the co-evolution (Silk \& Rees 1998, King 2004).
However there is increasing evidence for a delayed suppression of the star formation (Rafferty et al. 2006). More 
sophisticated observations are needed for future test of the coevolution.

As we show the gas-to-dust ratios in Seyfert 2 galaxies may be different from each other. Future identification 
of low$-{\cal Q}$ unabsorbed Seyfert 2 galaxies by polarization 
will be interesting to fill all members in the Seyfert family. Also a cross identification of {\em ROSAT} with
SDSS type II AGN will be interesting since this will provide a large sample of the unabsorbed type II
AGNs. Such a large sample will shed light of the nature of the unabsorbed type II AGNs, such as the evolving 
opening angle of the torus, the star formation history and the feedback processes.

Finally we have to point out a possible kind of Seyfert 2 galaxies in Figure 1. For low ${\cal Q}$ Seyfert
2s, there is a possibility that the black hole mass is less massive 
($\sim 10^6\sunm$ like NLS1s) and have no polarized BLR region. Such a kind of 
Seyfert 2 galaxies is characterized by: 1) it has no broad Balmer emission 
lines, but has "narrower" broad line region like absorbed HBLR S2s; 2) hydrogen column density is low, 
so soft X-ray are relatively bright; 3) it has no polarized broad lines. The objects are 
evolving into high$-{\cal Q}$ Seyfert 2s as shown in Fig 1.
Such an object may be formed in very strong interaction between galaxies, the star burst is so intensive 
that the dust is produced in a short timescale compared with its age and thus
has a very dust-rich nuclear region. However we stress here this
is only an imagination. Future observations are expected to find such a kind of objects.

\section{Conclusions} 
We have systematically studied the distributions of the black holes masses and accretion rates in
NLS1s, BLS1s, absorbed non-HBLR S2s, HBLR S2s and unabsorbed Seyfert 2s. We find a
sequence of the black hole masses from small to large: NLS1/absorbed non-HBLR S2s $\rightarrow$
BLS1s/HBLR S2s $\rightarrow$ unabsorbed Seyfert 2 galaxies. A queue of the Eddington ratios from large
to small is also found: NLS1/absorbed non-HBLR S2s $\rightarrow$
BLS1s/HBLR S2s $\rightarrow$ unabsorbed non-BLR Seyfert 2s. The two queues imply an evolutionary sequence
of Seyfert galaxy nuclei due to accretion.
We also find that the opening angle of the torus is evolving with the growth of the
black holes. Feedback from the black hole activities is evidenced from the deficiency of dust in absorbed
Seyfert 2 galaxies compared with the prediction of Schmidt's law. We obtain the lifetimes of NLS1s and 
BLS1s from the sample according to their appearance frequencies. The orientation-based unification should be
improved by including other four parameters:  black hole mass, accretion rates, changing opening angle, 
and gas-dust ratio (${\cal Q}$) in order to unify {\em all } kinds of Seyfert galaxies.

There are many things to be examined in future investigations. Future spectropolarmetric observations are
necessary to test the existence of HBLR in unabsorbed Seyfert 2 galaxies (they are too limited in the present sample).
Second the radio properties of the unabsorbed Seyfert 2 galaxies 
will further enhance the current understanding of their nature. Particularly the radio-loudness will discover 
their Eddington ratio according to an anti-correlation between the radio-loudness and the Eddington ratio 
(Ho 2000). Third spectropolarimetric observations of soft X-ray-selected unabsorbed Seyfert 2 galaxies are 
needed in order to test whether they contain hidden broad line regions. This may enhance the present 
classification of unabsorbed Seyfert 2 galaxies. It is urgent to issue the formation of the reflecting mirror 
in Seyfert 2s for a more detailed evolutionary sequence. 
Theoretical work on the polarization due to various 
kinds of dust grains in different geometry will probe the gas-to-dust ratio and unambiguously reveal the nature 
of the dusty torus in Seyfert 2 galaxies (Onaka 1995, Wolf, et al. 2002). 
Fourth, the black hole masses in NLS1s are still a matter of debate (Ryan et al. 2006). Some NLS1s have larger 
black hole masses. This may be either incorrectly estimated or those objects are not real narrow line objects.
It is very important for future work to assemble "pure" NLS1s is very important to estimate their mass via 
methods independent to reverberation mapping or empirical relation 
to test the proposed evolutionary scenario. Finally, the roles of very low luminosity AGNs from Hao et al. 
sample in the unification scheme is poorly understood. Extension to this sample will extend and enhance
the results in this paper.

\acknowledgements{The authors are grateful to the anonymous referee for a large number of comments and suggestions
improving the manuscript. L. C. Ho and S. Komossa are acknowledged for useful comments and simulating discussions. 
Special thanks 
are given to Y.-M. Chen and W.-H Bian for checking data and interesting discussions. We highly appreciate the
useful discussions among members in the IHEP (Institute of High Energy Physics) AGN group.
This research is supported by Natural Science Foundation of China through NSFC-10325313, 
NSFC-10233030 and NSFC-10521001.}

\clearpage
{\rotate
\begin{deluxetable}{llll}

\tabletypesize{\footnotesize}
\tablewidth{0pt}
\tablecaption{{\sc The Cateogories of Seyfert Galaxies}
\label{tbl-1}}
\tablehead{
Class & abbreviation &observed properties & comments and notes \\ }                                                    
\startdata
Broad emission line Seyfert 1s      &BLS1s       &showing strong broad permitted emission lines,\\ 
                                    &            &FWHM(H$\beta)>$2000\kms\\ 
Narrow emission line Seyfert 1s     &NLS1s       &similar to BLS1 spectrum, but FWHM(H$\beta$)<2000\kms \\
                                    &            &and flux ratio of H$\beta/$\oiii$<3$ \\
Hidden broad line region Seyfert 2s &HBLR S2s    &their spectrum is dominated by starlight in the optical-UV \\ 
                                    &            &band, but for which but for which spectropolarimetry reveals\\
                                    &                       &hidden broad-lines in the polarized light ("hidden broad-line\\
                                    &                       &region")\\
X-ray absorbed HBLR S2s             &absorbed HBLR S2s      &HBLR S2 spectrum, but $N_{\rm H}\ge 10^{22}$cm$^{-2}$\\
X-ray unabsorbed HBLR S2s           &unabsorbed HBLR S2s    &HBLR S2 spectrum, but $N_{\rm H}\le 10^{22}$cm$^{-2}$&PB02, M02, B03, M05, G06\\
Non-HBLR S2s                        &non-HBLR S2s           &absence of the permitted broad emission lines in polarized light&K94, T03 \\        	        
X-ray absorbed non-HBLR S2s         &absorbed non-HBLR S2s  &optical non-HBLR S2s, and $N_{\rm H}\ge 10^{22}$cm$^{-2}$\\
X-ray unabsorbed non-HBLR S2sA     & absorbed non-HBLR S2sA &optical non-HBLR S2s, but $N_{\rm H}\le 10^{22}$cm$^{-2}$, less massive BH& predicted in this paper\\
X-ray unabsorbed non-HBLR S2sB     &unabsorbed non-HBLR S2sB&optical non-HBLR S2s, but $N_{\rm H}\le 10^{22}$cm$^{-2}$, more massive BH&could be final stage
\enddata					        	        	
\tablecomments{B03: Boller et al. (2003); G06: Gallo et al. (2006); K94: Kay (1994); M02: Mainieri et al. (2002);
M05: Mateos et al. (2005); PB02: Panessa \& Bassini (2002); T03: Tran (2003)}   
\end{deluxetable}
}

\clearpage
\pagestyle{empty}
{\rotate
\begin{deluxetable}{lccccccl}

\tabletypesize{\footnotesize}
\tablewidth{0pt}
\tablecaption{{\sc The List of Seyfert Galaxy Samples }
\label{tbl-1}}
\tablehead{
Sample                    & $z$ & $N_{\rm total}$ & $N_{\rm NLS1}$  & $N_{\rm BLS1}$  & $N_{\rm non-HBLS2}$ &  $N_{\rm HBLS2}$ & notes \\
                          &    &               &       &       & (absorbed)&        &       \\ 
   (1)  & (2)  &(3)  &   (4) &  (5)    & (6)        &  (7)          &  (8)          }
\startdata
de Grijp et al.     1987 & $0.000\sim0.914$ & 306 & 14  & 109 & 25 & 22   & 133 S2s without polarimetric observation; 3 S2 are unabsorbed NHBLR S2\\
 Edelson             1987 & $0.002\sim0.061$ & 42  & 5   & 18  & 6   & 6   & 7 S2s without polarimetric observation\\ 
 Low et al.          1988 & $0.008\sim0.163$ & 27  & 6   & 21  & ... & ... & \\
 Osterbrock \& Shaw  1988 & $0.002\sim0.063$ & 37  & 1   & 2   & ... & ... & 20 HII; 3 LINER 11 S2s without polarimetric observation\\
 Sandrs et al.       1988 & $0.011\sim0.070$ & 9   & ... & 1   & ... & ... & 8 S2s without polarimetric observation\\ 
 Salzer              1989 & $0.012\sim0.019$ & 17  & ... & 9   & 1   & ... &   7 S2s without polarimetric observation\\
 Spinoglio \& Malkan 1989 & $0.002\sim0.172$ & 58  & 7   & 19  & 10  & 19  & 13 S2s without polarimetric observation; 1 S2 is unabsorbed NHBLR S2\\ 
 Miller \& Goodrich  1990 & $0.012\sim0.051$ & 8   & ... & ... & 4   & 4   & \\  
 Huchra \& Burg      1992 & $0.000\sim0.158$ & 85  & 4   & 23  &  10  &  5  & 32 LINER; 2 X-ray source; 1 BL Lac object; 7 S2s without polarimetric observation \\
 Wilson \& Tsvetanov 1994 & $0.003\sim0.037$ & 11  & ... & 2   &  4  & 4   & 1 S2 without polarimetric observation\\
 Maoilino \& Rieke   1995 & $0.000\sim0.034$ & 91  & 6   & 13  & 24  & 11  & 37 S2s without polarimetric observation\\
 Young et al.        1996 & $0.011\sim0.181$ & 24  & ... & ... & 11  & 13  & according their Table 4, 5\\
 Heisler et al.      1997 & $0.003\sim0.062$ & 16  & ... & ... & 9   & 7   & \\ 
 Grupe et al.        1999 & $0.021\sim0.545$ & 76  & 26  & 48  & ... & ... & 2 objects without $\rm FWHM_{\rm H \beta}$ \\
 Moran               2000 & $0.001\sim0.015$ & 31  & ... & ... & 20  & 11  & \\   
 Schmitt\& Kinney    2000 & $0.002\sim0.051$ & 141 & 9   & 39  & 18  & 26  & 49 S2s without polarimetric observation \\ 
 Alexander           2001 & $0.003\sim0.062$ & 16  & ... & ... & 9   & 7   & \\
 Lumsden et al.      2001 & $0.001\sim0.073$ & 28  & ... & ... & 9   & 8   & 11 objects uncertain for S/N ratio\\ 
 Nagao et al.        2001 & $0.002\sim0.632$ & 119 & 36  & 83  & ... & ... & \\
 Veron-Cetty et al.  2001 & $0.002\sim0.100$ & 83  & 64  & 19  & ... & ... & \\
 Gu \& Huang         2002 & $0.001\sim0.181$ & 90  & ... & ... & 49  & 41  & \\
 Crenshaw et al.     2003 & $0.002\sim0.084$ & 97  & 19  & 78  & ... & ... & \\
 Marziani et al.     2003 & $0.001\sim0.773$ & 215 & 23  & 192 & ... & ... & 41 BL PG-QSO; 8 NL PG-QSO; 23 BL 3C galaxy\\  
 Tran                2003 & $0.001\sim0.073$ & 112 &  6  & 40  & 27  & 22  & 17 LINERs\\
 Deluit              2004 & $0.005\sim0.173$ & 16  & ... & ... & 8   & 8   & with {\em Beppo}SAX observations\\ 
 Grupe et al.        2004 & $0.002\sim0.349$ & 110 & 50  & 60  & ... & ... & \\ 		        	        
\enddata					        	        	
\tablecomments{The reference queue is ranked by the publishing year.}   
\end{deluxetable}
}

\clearpage

\begin{deluxetable}{lcccccccccccl}
\rotate

\tabletypesize{\footnotesize}
\tablewidth{0pt}
\tablecaption{{\sc The Broad Line Seyfert 1 Sample}
\label{tbl-1}}
\tablehead{
  Name &$z$& morphology & $M_{\rm bul}$ & FWHM     &$\log M_{\rm BH}$ &$\log \cal E$&FWHM &$\log M_{\rm BH}$&$\log \cal E$&$\log L_{\rm [O~ III]}$&$\Gamma_{\rm HX}$&Ref. \\
       &   &            &               &(H$\beta$)&($\sunm$)         &             &(\oiii)&($\sunm$)      &             & (\ergs)               &                 &        \\  
   (1) &(2)&    (3)     &    (4)        &  (5)     &        (6)       &     (7)     & (8) &        (9)      &    (10)     &    (11)               &    (12)         &  (13) }
                                                   
\startdata
 1ES0459+034              & 0.016 & E                    & $-19.29$ &  4320 &  7.24 &  $-1.52$ & 360 &  7.15 &  $-0.18 $ & 41.57 & $ ... $               & 43, 23, 23, 23     \\ 
 1H1142-178               & 0.033 & Irr                  & $  ... $ &  3555 &  7.01 &  $-0.15$ & ... &   ... &  $  ... $ & 41.90 & $ ... $               &      1,      1     \\ 
 1H1927-516               & 0.040 & (R)SB0/a             & $-19.98$ &  3814 &  7.18 &  $-0.51$ & ... &   ... &  $  ... $ & 41.58 & $ ... $               & 43,  1,      1     \\ 
 1H1934-063               & 0.011 & ...                  & $  ... $ &  5531 &  7.32 &  $-1.84$ & 541 &  7.86 &  $-1.16 $ & 41.30 & $ ... $               &     19, 19, 24     \\
 1H2107-097               & 0.027 & ...                  & $  ... $ &  2024 &  7.15 &  $-0.51$ & ... &   ... &  $  ... $ & 42.14 & $ ... $               &      1,      1     \\ 
 2E1615+0611              & 0.038 & ...                  & $  ... $ &  3851 &  7.15 &  $-0.72$ & ... &   ... &  $  ... $ & 41.99 & $ ... $               &      1,      1     \\ 
 3A0557-383               & 0.034 & ...                  & $  ... $ &  2500 &  7.64 &  $-0.36$ & ... &   ... &  $  ... $ & 42.03 & $ ... $               &     20,     26     \\
 4U0241+61                & 0.045 & ...                  & $  ... $ & 14071 &  8.34 &  $-0.36$ & ... &   ... &  $  ... $ & 41.29 & $ ... $               &      1,      1     \\ 
 AKN120                   & 0.033 & Sb/pec               & $-19.98$ &  6128 &  8.25 &  $-1.23$ & 490 &  7.69 &  $-0.24 $ & 42.05 & $1.91\pm0.03$         & 43,  1,  3,  1, 49 \\ 
 AM2354-304               & 0.033 & ...                  & $  ... $ &  2400 &  6.89 &  $-0.89$ & 250 &  6.52 &  $-0.45 $ & 40.66 & $ ... $               &      2,  3,  2     \\ 
 B09.01                   & 0.043 & ...                  & $  ... $ &  3283 &  6.89 &  $-0.95$ & ... &   ... &  $  ... $ & 41.57 & $ ... $               &      1,      1     \\ 
 B25.02                   & 0.043 & ...                  & $  ... $ &  5345 &  7.75 &  $-0.86$ & ... &   ... &  $  ... $ & 41.89 & $ ... $               &      1,      1     \\ 
 CASG855                  & 0.040 & Spiral               & $-18.53$ &  4040 &  7.24 &  $-1.42$ & 200 &  6.13 &  $ 0.74 $ & 41.46 & $ ... $               & 43,  2,  2,  2     \\ 
 ESO141-G55$^{\ast}$      & 0.037 & S?                   & $-19.01$ &  5326 &  7.91 &  $-0.67$ & 350 &  7.10 &  $ 0.40 $ & 42.10 & $1.72\pm0.06$         & 43,  1,  3,  1, 52 \\ 
 ESO198-G024              & 0.046 & ...                  & $  ... $ &  6400 &  8.28 &  $-1.32$ & ... &   ... &  $  ... $ & 42.13 & $1.78 $               &     20,     20     \\
 ESO323-G77               & 0.015 & (R)SB(l)0            & $-19.32$ &  2500 &  7.39 &  $-0.56$ & ... &   ... &  $  ... $ & 41.68 & $ ... $               & 43, 20,     20     \\
 ESO344-G016              & 0.040 & Sb                   & $-19.98$ &  2500 &  7.45 &  $-0.51$ & ... &   ... &  $  ... $ & 41.77 & $ ... $               & 43, 17,     20     \\
 ESO362-G018$^{\ast}$     & 0.013 & S0/a                 & $-18.92$ &  4000 &  7.49 &  $-1.21$ & ... &   ... &  $  ... $ & 41.72 & $ ... $               & 43, 17,     15     \\
 ESO383-G035$^{\ast}$     & 0.008 & E-S0                 & $-18.36$ &  2400 &  7.23 &  $-0.62$ & ... &   ... &  $  ... $ & 39.45 & $1.80 $               & 43, 20,     26, 50 \\
 ESO438-G009              & 0.023 & (R'$_{-}1$)SB(rl)ab  & $-19.38$ &  5000 &  7.78 &  $-1.33$ & 250 &  6.52 &  $ 0.51 $ & 41.62 & $ ... $               & 43, 17, 25, 25     \\
 EXO1128+69               & 0.045 & ...                  & $  ... $ &  2130 &  6.98 &  $-0.63$ & 200 &  6.13 &  $ 1.23 $ & 41.95 & $ ... $               &      2,  2,  2     \\ 
 F1146                    & 0.032 & Sb                   & $-17.91$ &  4300 &  7.83 &  $-1.05$ & ... &   ... &  $  ... $ &  ...  & $ ... $               & 43, 17             \\
 F265                     & 0.030 & Sa                   & $-17.90$ &  2800 &  7.33 &  $-0.78$ & ... &   ... &  $  ... $ & 41.73 & $ ... $               & 43, 20,     20     \\
 F51                      & 0.014 & (R'$_{-}2$)SB(rs)b   & $-17.51$ &  3079 &  6.84 &  $-0.65$ & 210 &  6.21 &  $ 0.26 $ & 41.07 & $ ... $               & 43,  1,  3,  1     \\ 
 F9                       & 0.046 & S                    & $-22.24$ &  6270 &  8.05 &  $-0.70$ & 425 &  7.44 &  $-0.09 $ & 41.95 & $1.82\pm0.06$         & 43,  1,  3,  1, 49 \\ 
 H0307-730                & 0.028 & Sa                   & $-19.35$ &  2900 &  7.35 &  $-0.82$ & ... &   ... &  $  ... $ & 41.31 & $ ... $               & 43, 20,     27     \\
 IC4329A$^{\ast}$         & 0.016 & SA0+: sp             & $-19.43$ &  4800 &  7.42 &  $-0.95$ & 550 &  7.89 &  $-0.94 $ & 41.55 & $1.71 $               & 43, 20,  3,  1, 49 \\ 
 IISZ10                   & 0.034 & ...                  & $  ... $ &  3765 &  7.34 &  $-0.82$ & ... &   ... &  $  ... $ & 41.55 & $ ... $               &      1,      1     \\
 IRAS04448-0513           & 0.044 & ...                  & $  ... $ &  2699 &  7.48 &  $-0.61$ & ... &   ... &  $  ... $ & 41.64 & $ ... $               &     19,     15     \\
 IRAS05078+1626           & 0.018 & ...                  & $  ... $ &  3904 &  7.20 &  $-1.40$ & 308 &  6.88 &  $ 0.53 $ & 42.01 & $ ... $               &     19, 19, 19     \\
 IRAS17216+3633           & 0.040 & ...                  & $  ... $ &  4837 &  7.69 &  $-1.34$ & 540 &  7.86 &  $-0.29 $ & 42.17 & $ ... $               &     19, 19, 15     \\
 IRAS23226-3843           & 0.036 & ...                  & $  ... $ &  9500 &  8.22 &  $-1.98$ & 320 &  6.95 &  $-0.36 $ & 41.18 & $ ... $               &      2,  2,  2     \\ 
 IRASF05563-3820$^{\ast}$ & 0.034 & ...	                 & $  ... $ &  2500 &  7.63 &  $-0.36$ & ... &  ...  &  $  ... $ & 41.98 & $ ... $               &     20, 19, 20     \\ 
 MCG+8-11-11              & 0.020 & SB0                  & $-18.91$ &  3630 &  7.89 &  $-0.75$ & 605 &  8.06 &  $-0.45 $ & 42.21 & $1.56\pm0.08$         & 43, 17,  3,  4, 49 \\
 MCG+8-15-56              & 0.041 & Spiral               & $-20.08$ &  2247 &  7.22 &  $-0.53$ & 571 &  7.96 &  $-0.74 $ & 41.81 & $ ... $               & 43, 19, 19, 19     \\
 MS04124-0802             & 0.037 & SO                   & $-19.60$ &  6338 &  7.89 &  $-1.17$ & ... &   ... &  $  ... $ & 42.85 & $ ... $               & 43,  1,      1     \\ 
 Mark10                   & 0.030 & SBbc                 & $-19.73$ &  2400 &  7.14 &  $-0.69$ & 360 &  7.15 &  $-0.75 $ & 41.00 & $ ... $               &  3, 17,  4,  4     \\
 Mark1040$^{\ast}$        & 0.016 & Sbc                  & $-17.61$ &  4220 &  7.92 &  $-0.96$ & 304 &  6.86 &  $-0.14 $ & 41.31 & $1.56\pm0.14$         & 43, 18,  4,  4, 49 \\
 Mark1126                 & 0.010 & (R)SB(r)a            & $-18.00$ &  3142 &  6.90 &  $-1.29$ & 220 &  6.29 &  $ 0.29 $ & 41.18 & $ ... $               & 43, 18,  4,  3     \\
 Mark1146                 & 0.039 & Sab                  & $-19.65$ &  3142 &  7.58 &  $-0.76$ & ... &   ... &  $  ... $ & 41.39 & $ ... $               & 43, 18,      4     \\
 Mark1243                 & 0.035 & Sa                   & $-20.41$ &  3172 &  7.46 &  $-0.87$ & ... &   ... &  $  ... $ & 41.20 & $ ... $               & 43, 18,      4     \\
 Mark1310                 & 0.019 & E                    & $-19.92$ &  3000 &  6.77 &  $-1.33$ & ... &   ... &  $  ... $ & 41.37 & $ ... $               & 43,  2,      2     \\ 
 Mark141                  & 0.042 & E                    & $  ... $ &  3600 &  7.47 &  $-1.06$ & 330 &  7.00 &  $ 0.20 $ & 41.80 & $ ... $               &      2,  3,  2     \\ 
 Mark231$^{\ast}$         & 0.041 & SA(rs)c? pec         & $-19.15$ &  6418 &  8.33 &  $-1.08$ & ... &   ... &  $  ... $ & 42.73 & $1.89^{+1.09}_{-0.56}$& 43,  1,      1, 54 \\ 
 Mark236                  & 0.050 & SB                   & $-20.02$ &  4900 &  7.86 &  $-1.24$ & ... &   ... &  $  ... $ & 41.86 & $ ... $               & 43, 21,     26     \\
 Mark279                  & 0.031 & S0                   & $-20.30$ &  5425 &  7.92 &  $-1.19$ & 580 &  7.99 &  $-1.04 $ & 41.54 & $2.04\pm0.03$         & 43,  1,  4,  1, 49 \\ 
 Mark290                  & 0.030 & E1?                  & $-20.20$ &  5221 &  7.65 &  $-1.15$ & 380 &  7.25 &  $ 0.34 $ & 42.18 & $1.68\pm0.03$         & 43,  1,  3,  1, 49 \\ 
 Mark315                  & 0.040 & S0/a pec?            & $-20.07$ &  3700 &  7.73 &  $-0.90$ & 500 &  7.73 &  $-0.18 $ & 42.14 & $ ... $               &  3, 19,  4, 15     \\
 Mark316                  & 0.041 & E                    & $-21.01$ &  4689 &  7.86 &  $-1.17$ & 608 &  8.07 &  $-0.55 $ & 42.11 & $ ... $               & 43, 19, 19, 19     \\
 Mark352                  & 0.015 & SA0                  & $-18.49$ &  3800 &  7.19 &  $-1.36$ & 270 &  6.65 &  $-0.21 $ & 41.04 & $ ... $               & 43, 17,  3,  3     \\
 Mark358                  & 0.045 & SAB(rs)bc:           & $-19.50$ &  2235 &  7.20 &  $-0.54$ & 320 &  6.95 &  $-0.04 $ & 41.67 & $ ... $               &  3, 18,  3,  3     \\
 Mark372                  & 0.031 & S0/a                 & $-19.73$ &  5500 &  8.06 &  $-1.25$ & 400 &  7.34 &  $-0.43 $ & 41.50 & $ ... $               & 43, 17,  6, 28     \\
 Mark40                   & 0.020 & S0 pec               & $-17.81$ &  4042 &  7.47 &  $-1.24$ & 220 &  6.29 &  $ 0.34 $ & 41.23 & $ ... $               & 43, 18,  3,  3     \\
 Mark486                  & 0.038 & SBb?                 & $-19.19$ &  2073 &  6.94 &  $-0.15$ & 350 &  7.10 &  $-0.11 $ & 41.59 & $2.02^{+0.92}_{-0.95}$& 43,  1,  3,  1, 57 \\ 
 Mark504                  & 0.036 & SB0                  & $-18.20$ &  2300 &  7.04 &  $-0.70$ & 170 &  5.84 &  $ 0.79 $ & 41.23 & $ ... $               & 43, 21,  4,  4     \\
 Mark506                  & 0.043 & SAB(r)a              & $-20.07$ &  5647 &  7.59 &  $-0.66$ & 430 &  7.46 &  $-0.68 $ & 41.38 & $ ... $               & 43,  1,  3,  1     \\ 
 Mark509$^{\ast}$         & 0.035 & Compact              & $  ... $ &  3431 &  7.66 &  $-0.27$ & 520 &  7.80 &  $-0.04 $ & 42.37 & $ 1.91$               &      1,  3,  1, 50 \\ 
 Mark530$^{\ast}$         & 0.029 & SA(rs)b: pec         & $-19.76$ &  6560 &  8.60 &  $-1.10$ & 330 &  7.00 &  $-0.21 $ & 41.39 & $ ... $               & 43, 17,  4,  4     \\
 Mark541                  & 0.039 & Compact              & $  ... $ &  3300 &  7.96 &  $-0.54$ & ... &   ... &  $  ... $ & 41.54 & $ ... $               &     21,     26     \\
 Mark543                  & 0.026 & cIm?                 & $  ... $ &  3828 &  7.57 &  $-1.07$ & 257 &  6.57 &  $ 0.26 $ & 41.42 & $ ... $               &     19, 19,  4     \\
 Mark590                  & 0.027 & SA(s)a:              & $-20.32$ &  2635 &  7.38 &  $-0.50$ & 450 &  7.34 &  $-0.56 $ & 41.58 & $1.66\pm0.03$         & 43,  1,  5,  1, 53 \\ 
 Mark595                  & 0.028 & Sa                   & $-19.51$ &  2360 &  7.14 &  $-0.67$ & 410 &  7.38 &  $-0.20 $ & 41.78 & $ ... $               & 43, 17,  3, 29     \\
 Mark6$^{\ast}$           & 0.019 & SAB0+:               & $-18.96$ &  ...  & ...   &  $ ... $ & 690 &  8.29 &  $-0.57 $ & 42.32 & $1.81^{+0.22}_{-0.20}$&  3,   ,  5,  4, 55 \\ 
 Mark618$^{\ast}$         & 0.035 & SB(s)b pec           & $-19.71$ &  3018 &  7.15 &  $-0.32$ & 645 &  8.17 &  $-1.34 $ & 41.43 & $1.80^{+0.70}_{-0.50}$& 43,  1,  3,  1, 58 \\ 
 Mark704$^{\ast}$         & 0.029 & SBa   	         & $-20.12$ &  5684 &  7.76 &  $-0.97$ & 400 &  7.34 &  $-0.82 $ & 41.11 & $ ... $               &  3,  1,      1     \\
 Mark732                  & 0.029 & E+ pec               & $-21.24$ &  4428 &  7.87 &  $-1.07$ & 687 &  8.28 &  $-0.84 $ & 42.04 & $ ... $               & 43, 19, 19, 19     \\
 Mark79$^{\ast}$          & 0.022 & SBb                  & $-19.29$ &  4950 &  7.98 &  $-1.16$ & 390 &  7.29 &  $-0.33 $ & 41.56 & $1.82\pm0.03$         & 43, 17,  5, 28, 53 \\
 Mark817$^{\ast}$         & 0.033 & SBc                  & $-18.58$ &  4672 &  7.88 &  $-0.82$ & 663 &  7.00 &  $-0.90 $ & 41.92 & $ ... $               & 43,  1, 19,  1     \\ 
 Mark841                  & 0.036 & E                    & $-21.81$ &  5240 &  7.78 &  $-0.85$ & 774 &  6.52 &  $ 0.13 $ & 43.07 & $1.77 $               & 43,  1, 19,  1, 50 \\ 
 Mark871                  & 0.034 & SB0                  & $-19.87$ &  3691 &  7.25 &  $-0.61$ & 328 &  6.71 &  $ 0.09 $ & 41.68 & $ ... $               & 43,  1, 19,  1     \\ 
 Mark9$^{\ast}$           & 0.039 & S0 pec?              & $-20.57$ &  2900 &  7.63 &  $-0.60$ & 430 &  7.46 &  $-0.43 $ & 41.63 & $ ... $               & 43, 22,  3,  4     \\
 Mark926                  & 0.047 & ...                  & $  ... $ &  8467 &  8.33 &  $-0.96$ & 358 &  7.14 &  $ 0.54 $ & 42.28 & $1.73\pm0.04$         &      1,  4,  1, 49 \\ 
 Mark975                  & 0.050 & S?                   & $-20.76$ &  4123 &  7.42 &  $-0.47$ & 510 &  7.76 &  $-0.20 $ & 42.16 & $ ...$                & 43,  1,  4,  1     \\ 
 NGC1019                  & 0.024 & SB(rs)bc             & $-18.62$ &  4915 &  7.69 &  $-1.37$ & 372 &  7.21 &  $-0.58 $ & 41.23 & $ ...$                & 43, 19, 19, 19     \\
 NGC1566$^{\ast}$         & 0.004 & (R'$_{-}1$)SAB(rs)bc & $-18.72$ &  2583 &  6.11 &  $-1.61$ & 394 &  7.31 &  $-1.67 $ & 40.24 & $ ...$                & 43, 17, 33, 33     \\   
 NGC3227$^{\ast}$         & 0.003 & SAB(s) pec           & $-18.28$ &  3920 &  6.67 &  $-1.82$ & 514 &  7.78 &  $-1.56 $ & 40.81 & $1.56\pm0.1 $         & 43, 17,  4,  4, 47 \\
 NGC3516$^{\ast}$         & 0.009 & (R)SB(s)0:           & $-19.67$ &  4760 &  6.98 &  $-1.88$ & 248 &  6.50 &  $ 0.23 $ & 41.33 & $1.83\pm0.04$         & 43, 17,  4,  4, 49 \\
 NGC3783                  & 0.009 & (R')SB(r)a           & $-19.13$ &  3567 &  7.08 &  $-1.00$ & 230 &  6.37 &  $ 0.43 $ & 41.40 & $1.70\pm0.10$         & 43,  1,  3,  1, 49 \\ 
 NGC4151$^{\ast}$         & 0.003 & (R')SAB(rs)ab:       & $-17.66$ &  6437 &  7.31 &  $-1.57$ & 425 &  7.44 &  $-0.33 $ & 41.71 & $1.6\pm0.2  $         & 43,  1,  3,  1, 47 \\ 
 NGC4235                  & 0.008 & SA(s)a               & $-18.89$ &  7600 &  7.88 &  $-1.90$ & 400 &  7.34 &  $-1.84 $ & 40.09 & $ ... $               & 43, 17,  4, 30     \\
 NGC4593$^{\ast}$         & 0.009 & (R)SB(rs)b           & $-19.57$ &  5320 &  7.22 &  $-1.06$ & 255 &  6.55 &  $-0.62 $ & 40.53 & $1.78\pm0.05$         & 43,  1,  3,  1, 49 \\ 
 NGC526A$^{\ast}$         & 0.019 & S0 pec               & $-19.04$ &  ...  & ...   &   ...    & 210 &  6.21 &  $ 1.10 $ & 41.91 & $1.47 $               & 43,   ,  3,  3, 50 \\ 
 NGC5548$^{\ast}$         & 0.017 & (R')SA(s)0/a         & $-19.94$ &  4044 &  7.27 &  $-0.75$ & 410 &  7.38 &  $-0.35 $ & 41.63 & $1.68\pm0.05$         &  3, 44,  3,  1, 49 \\ 
 NGC5940                  & 0.033 & SBab                 & $-20.07$ &  5240 &  7.39 &  $-0.77$ & ... &   ... &  $  ... $ & 41.42 & $ ... $               & 43,  1,      1     \\ 
 NGC6212                  & 0.030 & Sb                   & $-18.80$ &  6050 &  7.95 &  $-1.49$ & 400 &  7.34 &  $-1.42 $ & 40.51 & $ ... $               & 43, 17,  3,  3     \\
 NGC6860$^{\ast}$         & 0.015 & (R')SB(r)ab          & $-18.98$ &  3900 &  7.59 &  $-1.09$ & ... &   ... &  $  ... $ & 41.17 & $ ... $               & 43, 17,     32     \\
 NGC7213$^{\ast}$         & 0.006 & SA(s)0               & $-20.28$ &  3200 &  7.19 &  $-1.09$ & 990 &  8.92 &  $-2.84 $ & 40.68 & $1.80 $               & 43, 17,  6, 15, 50 \\ 
 NGC7214                  & 0.023 & SB(s)bc pec:         & $-19.53$ &  4700 &  7.54 &  $-1.42$ & 530 &  7.83 &  $-0.99 $ & 41.43 & $ ... $               & 43,  2,  2,  2     \\ 
 NGC7469$^{\ast}$         & 0.017 & (R')SAB(rs)a         & $-20.16$ &  2648 &  7.23 &  $-0.43$ & 360 &  7.15 &  $ 1.25 $ & 43.00 & $2.01 $               & 43,  1,  3,  1, 50 \\ 
 NGC985                   & 0.043 & SBbc? p              & $-20.22$ &  7499 &  8.15 &  $-1.06$ & 320 &  6.95 &  $ 0.90 $ & 42.44 & $1.53 $               & 43,  1,  3,  1, 50 \\ 
 UGC00524                 & 0.036 & (R')SB(s)b           & $-19.87$ &  4161 &  7.80 &  $-1.02$ & ... &   ... &  $  ... $ &  ...  & $ ... $               & 43, 18             \\
 UGC03142                 & 0.022 & (R)S0/a              & $-17.38$ & 10489 &  8.29 &  $-2.08$ & ... &   ... &  $  ... $ &  ...  & $ ... $               & 43, 18             \\
 UGC03223                 & 0.016 & SBa                  & $-19.19$ &  4740 &  7.66 &  $-1.34$ & 255 &  6.55 &  $-0.19 $ & 40.96 & $ ... $               & 43, 17,  6, 26     \\
 UGC10683B                & 0.032 & ...                  & $  ... $ &  8957 &  8.22 &  $-1.89$ & 616 &  8.09 &  $-1.13 $ & 41.56 & $ ... $               &     19, 19, 31     \\          
\enddata					        	        									        
\vglue -0.7cm
\tablecomments {The objects marked by * are from 12$\mu$m sample, and the same for other tables.
(1) source name; (2) redshift; (3) Hubble type; (4) buge absolute magnitude in B; (5) FWHM of $ \rm H \beta $ line (in $\rm km~s^{-1}$); 
(6) FWHM of \oiii line (in $\rm km~s^{-1}$); (7) mass of black hole (calculated with $ \rm H \beta $);					        
(8) Eddington ratio ($L_{\rm bol}=9 \times L_{\rm 5100}$ \AA); (9) mass of black hole (calculated with \oiii); 				        
(10) Eddington ratio ($L_{\rm bol}=3500 \times L_{\rm [O~ III]}$); (11) luminosity of \oiii $\lambda 5007$ emission line;  			        
(12) hard X-ray (2-10 keV) photon index; (13) reference (for column (3),(4),(5),(8),(11),(12) respectively).} 				        
\tablerefs{(1)Marziani et al. 2003; (2)Grupe et al. 2003; (3) Whittle 1992; (4) Dahari 1988; (5) Feldman et al. 1982; 
(6) Nelson \& Whittle 1995; (7) Vader et al. 1993; (8) Lipari \& Pastoriza 1991; (9)Young et al. 1996; (10) Wilson \& Nath 1990; 
(11) Veron 1981; (12) Appenzeller \& Ostreicher 1988; (13) Gu \& Huang 2002; (14) Tran 2003; (15) de Grijp 1992; 
(16) Bassani et al. 1999; (17) Crenshaw et al. 2003; (18) Botte et al. 2004; (19) Wang et al. 2005; (20) Winkler 1992; 
(21) Osterbrock 1977; (22) Peterson et al. 1984; (23) Ghigo et al. 1982; (24) Rodroguez-Ardila et al. 2000; 
(25) Kollatschny \& Fricke 1983; (26) Bonatto \& Pastoriza 1997; (27) Remillard et al. 1986; (28) Adams \& Weedman 1975; 
(29) Stirpe 1990; (30) Ho et al. 1997; (31) Wilson et al. 1981; (32) Lipari et al. 1993; (33) Kriss et al. 1991; (34) Veilleux 1991 ; 
(35) Kim et al. 1995; (36) Goodrich 1992; (37) Whittle et al. 1988; (38) Osterbrock \& De Robertis 1985; (39) Shu et al. 2006; 
(40) Gu et al. 2006; (41) Lumsden et al. 2004; (42) Schmitt et al. 2003; (43) NED; (44) Zhou \& Wang 2004  
(45) Imanishi 2002; (46) Imanishi 2003; (47) Cappi et al. 2006; (48) Pappa et al. 2005; (49) Turner et al. 1999;
(50) Tartarus; (51) Guainazzi et al. 2005; (52) Gondoin 2004; (53) Gallo et al. 2006; (54) Turner  1999; (55) Immler, et al. 2003;
(56) Matt et al. 2003; (57) Gallagher et al. 2002; (58) Rao et al. 1992.}
\end{deluxetable}    

\clearpage
\begin{deluxetable}{lcccccccccccl}
\rotate
\tabletypesize{\footnotesize}
\tablewidth{0pt}
\tablecaption{{\sc The Absorbed Hidden BLR Seyfert 2 Sample}
\label{tbl-1}}
\tablehead{
 Name             &$z$     & morphology &$M_{\rm bul}$  &FWHM     &$\log\mbh$ &$\log {\cal E}$&$\log L_{\rm [O~III]}$ &$\log N_{\rm H}$&$\log L_{\rm PAH}$ & $\Gamma_{\rm HX}$ &  Ref. \\ 
                  &        &            &               &(\oiii)  &($\sunm$)  &               &(\ergs)                &(cm$^{-2}$)     &(\ergs)            &                   &       \\
 (1)              &(2)     &(3)         &   (4)         &    (5)  &   (6)     &  (7)          &    (8)                &    (9)         &       (10)        &      (11)         &(12) }    
\startdata
 Circinus                  & 0.001 & SA(s)b:             & $-14.37 $ & ... &  ...  & $  ...  $ & 40.47  & $24.63  $ &  ...  & $1.60^{+0.50}_{-0.40}$ & 43,     13, 39, 46, 16 \\ 
 ESO273-IG04               & 0.039 & spirals?            & $-20.25 $ & ... &  ...  & $  ...  $ & 42.37  & $ ...   $ &  ...  & $... $                 & 43,     13             \\
 ESO434-G40                & 0.008 & (RL)SA(l)0          & $-17.85 $ & 220 &  6.29 & $ -1.05 $ & 39.84  & $22.30  $ &  ...  & $1.95^{+0.10}_{-0.09}$ & 43, 10, 41, 39, 46, 16 \\
 F02581-1136               & 0.030 & SAB(rs)a pec:       & $-19.90 $ & ... &  ...  & $  ...  $ & 41.07  & $ ...   $ &  ...  & $... $                 & 43,     14             \\
 F04385-0828$^{\ast}$      & 0.015 & S0                  & $-19.29 $ & 907 &  8.77 & $ -2.40 $ & 40.97  & $ ...   $ & 40.26 & $... $                 & 43,  7, 38             \\
 F15480-0344$^{\ast}$      & 0.030 & S0                  & $-20.60 $ & 664 &  8.22 & $ -0.57 $ & 42.25  & $22.81  $ & 40.77 & $2.0 $                 & 43,  8, 15, 14, 46, 51 \\
 IC5063$^{\ast}$           & 0.011 & SA(s)0+:            & $-19.72 $ & 370 &  7.20 & $ -0.59 $ & 41.21  & $23.38  $ &  ...  & $1.80\pm0.20 $         & 43, 11, 40, 14         \\
 IRAS05189-2524$^{\ast}$   & 0.043 & pec                 & $  ...  $ & 540 &  7.86 & $ -0.68 $ & 41.78  & $22.76  $ &  ...  & $1.70$                 & 43, 35, 35, 39,     16 \\
 IRAS13197-1627            & 0.017 & SB?                 & $-18.52 $ & ... &  ...  & $  ...  $ & 42.32  & $23.88  $ & 40.26 & $... $                 & 43,     13, 14, 46     \\
 IRAS18325-5926            & 0.020 & Sa                  & $-20.31 $ & ... &  ...  & $  ...  $ & 42.76  & $22.16  $ &  ...  & $1.85\pm0.08 $         & 43,     16, 39,     39 \\
 MCG-3-58-7$^{\ast}$       & 0.025 & (R')SAB(s)0/a       & $-19.32 $ & ... &  ...  & $  ...  $ & 41.69  & $ ...   $ & 40.89 & $... $                 & 43,     13,     46     \\
 Mrk1210                   & 0.014 & Sa                  & $-18.39 $ & ... &  ...  & $  ...  $ & 42.20  & $>24.00 $ &  ...  & $0.90^{+1.00}_{-1.10}$ & 43,     13, 13,     16 \\
 Mrk3                      & 0.014 & S0                  & $-19.11 $ & 780 &  8.50 & $ -0.66 $ & 42.44  & $24.13  $ &  ...  & $1.56^{+0.14}_{-0.26}$ & 43, 10, 15, 39,     16 \\
 Mrk78                     & 0.037 & SB                  & $-19.62 $ & 581 &  7.99 & $ -0.61 $ & 41.98  & $ ...   $ & 41.03 & $... $                 & 43, 36, 37,     45     \\
 Mrk348$^{\ast}$           & 0.015 & SA(s)0/a            & $-19.14 $ & 365 &  7.18 & $ -0.58 $ & 41.20  & $23.04  $ & 40.58 & $1.59^{+0.15}_{-0.14}$ & 43,  3, 15, 13, 46, 16 \\
 Mrk463E                   & 0.050 & ...                 & $  ...  $ & ... &  ...  & $  ...  $ & 42.97  & $23.20  $ & 41.17 & $1.40\pm0.90 $         & 43,     14, 14, 45, 16 \\
 Mrk477                    & 0.038 & Compact             & $  ...  $ & 370 &  7.20 & $  1.01 $ & 42.81  & $>24.00 $ & 41.18 & $0.2^{+0.80}_{-0.70}$  & 43,  3,  3, 39, 45, 16 \\
 NGC1068$^{\ast}$          & 0.004 & (R)SA(rs)b          & $-19.87 $ & 822 &  8.60 & $ -0.95 $ & 42.25  & $25.00  $ & 40.43 & $1.0\pm0.1 $           & 43, 34,  3, 14, 46,  4 \\
 NGC2110                   & 0.008 & SAB0-               & $-18.59 $ & 570 &  7.96 & $ -1.21 $ & 41.35  & $22.17  $ &  ...  & $1.36^{+0.07}_{-0.08}$ & 43,  4,  4, 39,     16 \\
 NGC2273                   & 0.006 & SB(r)a:             & $-18.34 $ & 165 &  5.79 & $  0.65 $ & 41.04  & $>24.13 $ &  ...  & $1.09^{+0.43}_{-0.32}$ & 43,  6,  3, 39,     48 \\
 NGC3081                   & 0.008 & (R$_{-}$1)SAB(r)0/a & $-18.82 $ & 220 &  6.29 & $  0.05 $ & 40.94  & $23.82  $ &  ...  & $1.70^{+0.26}_{-0.35}$ & 43, 11, 40, 13,     16 \\
 NGC424$^{\ast}$           & 0.012 & (R)SB(r)0/a         & $-18.80 $ & 515 &  7.78 & $ -0.83 $ & 41.46  & $24.00  $ &  ...  & $2.0 $                 & 43, 10, 13, 14,     56 \\
 NGC4388$^{\ast}$          & 0.008 & SA(s)b              & $-19.23 $ & 642 &  8.16 & $ -1.61 $ & 41.15  & $23.62  $ & 39.67 & $1.3\pm0.20 $          & 43,  9, 40, 14, 46, 47 \\
 NGC4507                   & 0.012 & SAB(s)ab            & $-19.26 $ & 237 &  6.42 & $  0.00 $ & 41.02  & $23.46  $ &  ...  & $1.61\pm0.20 $         & 43, 12,  4, 13,     16 \\
 NGC513$^{\ast}$           & 0.020 & Sb/c                & $-19.08 $ & 220 &  6.29 & $ -0.30 $ & 40.59  & $ ...   $ & 40.34 & $... $                 & 43,  6, 13    , 46     \\
 NGC5252                   & 0.023 & S0                  & $-20.11 $ & 454 &  7.56 & $ -0.56 $ & 41.60  & $22.63  $ & 40.75 & $1.45\pm0.20$          & 43,  6, 16, 13, 46, 16 \\
 NGC5506$^{\ast}$          & 0.006 & Sa pec              & $-18.10 $ & 430 &  7.46 & $ -0.45 $ & 41.61  & $22.53  $ &  ...  & $1.92^{+0.03}_{-0.02}$ & 43,  5, 13, 14,     16 \\
 NGC591                    & 0.015 & (R')SB0/a           & $-19.15 $ & 300 &  6.84 & $  0.24 $ & 41.68  & $>24.20 $ &  ...  & $... $                 & 43,  4,  3, 39         \\
 NGC6552                   & 0.027 & SB?                 & $-19.60 $ & ... &  ...  & $  ...  $ & 42.12  & $23.78  $ &  ...  & $1.4\pm0.40$           & 43,     13, 14,     16 \\
 NGC7212                   & 0.027 & Sab                 & $-18.85 $ & 435 &  7.48 & $ -0.61 $ & 41.47  & $23.65  $ &  ...  & $1.5^{+0.3}_{-0.6}$    & 43,  3, 42, 13,     51 \\
 NGC7314$^{\ast}$          & 0.005 & SAB(rs)bc           & $-17.92 $ & ... &  ...  & $  ...  $ & 41.88  & $22.02  $ &  ...  & $2.19^{+0.09}_{-0.06}$ & 43,     16, 39         \\     
 NGC7672                   & 0.013 & Sb                  & $-17.34 $ & ... &  ...  & $  ...  $ &  ...   & $25.00  $ &  ...  & $... $                 & 43,         13         \\
 NGC7674$^{\ast}$          & 0.029 & SA(r)bc pec         & $-19.45 $ & 350 &  7.10 & $  0.38 $ & 42.08  & $25.00  $ & 40.04 & $1.92\pm0.21 $         & 43,  3, 42, 14, 46, 16 \\
 NGC7682                   & 0.017 & SB(r)ab             & $-18.84 $ & 320 &  6.95 & $ -0.40 $ & 41.15  & $ ...   $ & 40.04 & $... $                 & 43,  5, 40,     46     \\
 NGC788                    & 0.014 & SA(s)0/a:           & $-19.89 $ & 190 &  6.04 & $ -0.46 $ & 40.18  & $23.32  $ &  ...  & $... $                 & 43,  3, 40, 39         \\
 Tol1238-364$^{\ast}$      & 0.011 & SB(rs)bc:           & $-18.25 $ & 300 &  6.84 & $ -0.04 $ & 41.40  & $25.00  $ & 40.05 & $... $                 & 43,  3,  3, 14, 45     \\ 	  
\enddata
\vglue -0.7cm					        
\tablecomments {(1) source name; (2) redshift; (3) Hubble type; (4) bulge absolute magnitude in B; (5) FWHM of \oiii line (in $\rm km~s^{-1}$); 
(6) mass of black hole; (7) Eddington ratio; (8) luminosity of \oiii $\lambda 5007$ emission line;
(9) column density; (10) luminosity of PAH; (11) hard X-ray (2-10 keV) photon index; (12) reference (for column (4),(5),(8),(9),(10),(11) respectively).}   

\tablerefs{(1)Marziani et al. 2003; (2)Grupe et al. 2003; (3) Whittle 1992; (4) Dahari 1988; (5) Feldman et al. 1982; 
(6) Nelson \& Whittle 1995; (7) Vader et al. 1993; (8) Lipari \& Pastoriza 1991; (9)Young et al. 1996; 
(10) Wilson \& Nath 1990; (11) Veron 1981; (12) Appenzeller \& Ostreicher 1988; (13) Gu \& Huang 2002; 
(14) Tran 2003; (15) de Grijp 1992; 
(16) Bassani et al. 1999; (17) Crenshaw et al. 2003; (18) Botte et al. 2004; (19) Wang et al. 2005; (20) Winkler 1992; 
(21) Osterbrock 1977; (22) Peterson et al. 1984; (23) Ghigo et al. 1982; (24) Rodroguez-Ardila et al. 2000; 
(25) Kollatschny \& Fricke 1983; (26) Bonatto \& Pastoriza 1997; (27) Remillard et al. 1986; (28) Adams \& Weedman 1975; 
(29) Stirpe 1990; (30) Ho et al. 1997; (31) Wilson et al. 1981; (32) Lipari et al. 1993; (33) Kriss et al. 1991;  
(34) Veilleux 1991 ; (35) Kim et al. 1995; (36) Goodrich 1992; (37) Whittle et al. 1988; 
(38) Osterbrock \& De Robertis 1985; (39) Shu et al. 2006; (40) Gu et al. 2006; (41) Lumsden et al. 2004;
(42) Schmitt et al. 2003; (43) NED; (44) Zhou \& Wang 2004; (45) Imanishi 2002; (46) Imanishi 2003;
(47) Cappi et al. 2006; (48) Pappa et al. 2005; (49) Turner et al. 1999;
(50) Tartarus; (51) Guainazzi et al. 2005; (52) Gondoin 2004; (53) Gallo et al. 2006; (54) Turner  1999; (55) Immler, et al. 2003;
(56) Matt et al. 2003; (57) Gallagher et al. 2002; (58) Rao et al. 1992; (59) Dewangan. \& Griffiths 2005.}
\end{deluxetable}    

\clearpage

\begin{deluxetable}{lccccccccccl}

\rotate
\tabletypesize{\footnotesize}
\tablewidth{0pt}
\tablecaption{{\sc The Unabsorbed S2 Sample}
\label{tbl-1}}
\tablehead{
 Name& $z$& morphology & $M_{\rm bul}$ &$\sigma$&FWHM &$\log M_{\rm BH}$&$\log{\cal E}$&$\log L_{\rm [O~III]}$& $N_{\rm H}$ &$\log L_{\rm PAH}$&Ref. \\ 
     &    &            &               &(\kms)  &(\oiii)&$(\sunm)$      &             & (\ergs)               &(cm$^{-2}$)  & (\ergs)          &        \\
(1)  &(2) &    (3)     &      (4)      &  (5)   & (6) &   (7)           &    (8)      &       (9)            &    (10)     &      (11)       &   (12) }
                                                   
\startdata
\multicolumn{12}{c}{ Unabsorbed HBLR S2}\\ \hline
F01475-0740$^{\ast}$(Pol)& 0.018 & E-S0            & $-18.80$ & ... & ... &$7.55^a$& $-0.46$  & 41.69 & $21.64$    & 40.34  & 1,        4, 25, 22 \\
IRAS00317-2142           & 0.027 &(R'$_-$1)SB(rl)bc& $-19.55$ & ... & 457 & 8.08   & $-1.55$  & 41.13 & $20.28$    &  ...   & 1,     6, 6,  5     \\ 
IRAS20051-1117           & 0.031 & ...             & $  ... $ & ... & 262 & 7.11   & $-0.23$  & 41.47 & $<21.60$   &  ...   &        6, 6,  5     \\ 
IC1631                   & 0.031 & Sab pec         & $-20.03$ & ... & 384 & 7.78   & $-0.39$  & 41.98 & $<21.50$   &  ...   & 1,    10, 5,  5     \\ 
NGC2992$^{\ast}$(Pol)    & 0.008 & Sa pec          & $-18.74$ & 158 & ... & 7.72   & $-0.96$  & 41.35 & $21.95$    &  ...   &23,  3,   24, 25     \\
NGC3185                  & 0.004 & (R)SB(r)a       & $-17.02$ &  61 & ... & 6.06   & $-0.80$  & 39.85 & $\le21.30$ &  ...   & 1,  3,   18, 12     \\ 
NGC5995$^{\ast}$(Pol)    & 0.025 & S(B)c           & $-17.97$ & ... & ... &$7.11^a$& $-0.25^b$& 41.86 & $21.93  $  &$<40.90$& 1,        5,  5, 22 \\\hline  
\multicolumn{12}{c}{ Unabsorbed non-HBLR S2}\\ \hline					    				   				   
IRAS01428-0404           & 0.018 & SB(rs)bc pec:   & $-18.35$ & ... & ... &$7.31^a$& $-2.59^b$&  ...  & $21.51$    &  ...   & 1,            5     \\ 
NGC1058                  & 0.002 & SA(rs)c         & $-15.16$ &  60 & ... & 6.03   & $-2.56$  & 38.06 & $\le21.78$ &  ...   & 1,  7,   18, 12     \\ 
NGC2685                  & 0.003 & (R)SB0+ pec     & $-17.66$ & 103 & ... & 6.97   & $-2.65$  & 38.92 & $\le21.48$ &  ...   & 1, 11,   18, 12     \\ 
NGC3031$^{\ast}$         & 0.001 & SA(s)ab         & $-18.89$ & 168 & ... & 7.83   & $-3.61$  & 38.81 & $\le21.00$ &  ...   & 1,  3,   18, 12     \\ 
NGC3147                  & 0.009 & SA(rs)bc        & $-19.39$ & 268 & ... & 8.64   & $-3.05$  & 40.19 & $<20.46$   &  ...   & 1,  7,   18,  5     \\ 
NGC3486                  & 0.002 & SAB(r)c         & $-15.93$ &  65 & ... & 6.17   & $-2.52$  & 38.25 & $\le21.48$ &  ...   & 1, 19,   18, 12     \\ 
NGC3660$^{\ast}$(Pol)    & 0.012 & SB(r)bc         & $-18.38$ & ... & ... &$7.33^a$& $-0.57$  & 41.36 & $20.26$    & 40.04  & 1,        4,  2, 22 \\ 
NGC3941                  & 0.003 & SB(s)0          & $-18.53$ & 131 & ... & 7.39   & $-3.18$  & 38.80 & $\le21.00$ &  ...   & 1, 13,   18, 12     \\ 
NGC4472                  & 0.003 & E2/S0           & $-21.03$ & 273 & ... & 8.67   & $-5.65$  & 37.62 & $21.48$    &  ...   & 1, 14,   18, 12     \\ 
NGC4501$^{\ast}$(Pol)    & 0.008 & SA(rs)b         & $-20.63$ & 151 & ... & 7.64   & $-2.40$  & 39.84 & $\le21.30$ &$<39.67$& 1, 15,   18, 12, 22 \\ 
NGC4565                  & 0.004 & SA(s)b? sp      & $-19.06$ & 144 & ... & 7.56   & $-2.79$  & 39.36 & $20.11$    &  ...   & 1, 17,   18,  5     \\ 
NGC4579$^{\ast}$         & 0.005 & SAB(rs)b        & $-19.49$ & 160 & ... & 7.74   & $-2.76$  & 39.58 & $20.39$    &  ...   & 1,  3.   18,  5     \\ 
NGC4594$^{\ast}$         & 0.004 & SA(s)a          & $-21.03$ & 250 & ... & 8.52   & $-3.86$  & 39.26 & $21.23$    &  ...   & 1,  8,   18,  5     \\ 
NGC4698                  & 0.003 & SA(s)ab         & $-17.70$ & 134 & ... & 7.43   & $-3.39$  & 38.64 & $20.91$    &  ...   & 1, 16,   18,  5     \\ 
NGC5033$^{\ast}$         & 0.003 & SA(s)c          & $-17.11$ & 138 & ... & 7.48   & $-2.61$  & 39.47 & $20.01$    &  ...   & 1, 15,   18,  5     \\ 
NGC5929$^{\ast}$(Pol)    & 0.008 & Sab: pec        & $-17.40$ & 121 & ... & 7.25   & $-1.41$  & 40.82 & $20.76$    &$<39.23$& 1,  3,    4,  2, 22 \\
NGC6251                  & 0.023 & E               & $-21.19$ & 293 & ... & 8.80   & $-4.33^b$& 41.77 & $21.88$    &  ...   & 1,        5,  5     \\
NGC676                   & 0.005 & S0/a: sp        & $-20.15$ & ... & ... &$8.27^a$& $-3.65$  & 39.21 & $\le21.00$ &  ...   & 1,       18, 12     \\  
NGC7590$^{\ast}$(Pol)    & 0.005 & S(r?)bc         & $-17.43$ & ... & ... &$6.83^a$& $-1.47$  & 39.95 & $<20.96$   &  ...   & 1,        2,  2     \\ 
NGC7679                  & 0.017 & SB0 pec:        & $-20.09$ & ... & 600 & 8.56   & $-2.81^b$& 41.78 & $20.34$    &  ...   & 1,    26, 5,  5     \\ 
 					        	        								      
\enddata
\tablecomments {(1) source name; (2) redshift; (3) Hubble type; (4) bulge absolute magnitude in B band;
(5) stellar velocity dispersion (in $\rm km~s^{-1}$); 
(6) FWHM of \oiii line (in $\rm km~s^{-1}$); (7): mass of black hole; 
(8) Eddington ratio; (9) luminosity of \oiii $\lambda 5007$ emission line;
(10) column density; (11) luminosity of PAH; (12) reference (for column (4),(5),(6),(7),(10),(11) respectively).
$^{\ast}$  12$\mu$m sample,
$^{a}$ $M_{\rm BH}$ is deduced from $M_{\rm bul}$.
$^{b}$ bolometric luminosity is deduced from $L_X$(2-10 KeV), we have
      IRAS01428-0404, $\log L_X=41.38$ (PB02);
      IRAS01428-0404, $\log L_X=41.38$ (PB02) 
      NGC5995,        $\log L_X=43.52$ (PB02); 
      NGC6251,        $\log L_X=41.38$ (PB02);
      NGC7679,        $\log L_X=42.42$ (Della Ceca et al. 2001)
}

\tablerefs{(1) NED; (2) Gu \& Huang 2002; (3) Nelson \& Whittle 1995; (4) Tran 2003; 
(5) Panessa \& Bassani 2002; (6) Moran et al. 1996; (7) McElroy 1995; (8) Kormendy 1988;
(9) Kirhakos \& Steiner 1990; (10) Prungniel \& Simien 1994; (11) Simien \& Prungniel 1997; 
(12) Cappi et al 2006; (13) Fisher 1997; (14) Merritt \& Ferrarese 2001; 
(15) Heraudeau \& Simien  1998; (16) Heraudeau et al. 1999; (17) de Souza et al. 1993; 
(18) Ho et al. 1997; (19) LEDA; (20) Beuing et al. 1999; (21) Wolter et al. 2006; 
(22) Imanishi 2003; (23) Whittle 1992; (24) Lumsden et al. 2004; (25) Shu et al. 2006; (26) Kim et al. 1995}

\end{deluxetable}

\clearpage
\begin{deluxetable}{lcccccccl}
\tabletypesize{\footnotesize}
\tablewidth{0pt}
\tablecaption{{\sc The Sample Completeness Analysis}
\label{tbl-1}}
\tablehead{
Sample             & $N_{\rm total}$ &$N_{\rm NLS1}$    &$N_{\rm BLS1}$   &  $N_{\rm HBLRS2}$ & $N_{\rm HBLRS2}$&$N_{\rm non-HBLRS2}$& $N_{\rm non-HBLRS2}$\\
                   &                 &                  &                 &    (absorbed)     &   (unabsorbed)  &    (absorbed)      &   (unabsorbed)        \\
  (1)              &       (2)       &       (3)        &      (4)        &        (5)        &      (6)        &      (7)           &      (8)               } 
\startdata
 Rush et al.(1993) &        83       &    9 ($\sim 11\%$)& 27 ($\sim 32\%$) & 20 ($\sim 24\%$)& 2 ($\sim 2\%$)  &  17 ($\sim 20\%$)  & 8 ($\sim 10\%$)      \\
 Tran       (2003) &         96      &    6 ($\sim 6\%$) & 40 ($\sim 42\%$) & 20 ($\sim 21\%$)& 3 ($\sim 3\%$)  &  24 ($\sim 25\%$)  & 3 ($\sim 3\%$)       \\
 our sample        &        243      &   44 ($\sim 18\%$)& 94 ($\sim 39\%$) & 36 ($\sim 15\%$)& 7 ($\sim 2.5\%$)  &  42 ($\sim 17\%$)  & 20 ($\sim 8\%$)      \\ 
mean fraction      &                 &$\sim 14\%$        &$\sim 38\%$       &$\sim 18\%$      &$\sim2.6\%$        & $\sim 20\%$        &$\sim 7\%$           \\          
\enddata
\vskip 0.1cm
\parbox{5.6in}
{\begin{itemize}
\item The mean fraction is obtained from the three samples. The percentage in the bracket indicates the relative number of eack 
sub-class Seyfert to the total.
\end{itemize}
} 
\end{deluxetable}


\begin{deluxetable}{lrrll}
\tabletypesize{\footnotesize}
\tablewidth{0pt}
\tablecaption{{\sc The Central Engines of BLS1s and HBLR S2s}
\label{tbl-1}}
\tablehead{
Parameters                               &    BLS1s      &    HBLR S2s    & $p_{\rm null}$ & Note       }
\startdata
$\langle \log \mbh\rangle$(H$\beta$)     &$7.53\pm0.05$  &$ ...$          &  ...           & isotropic  \\
$\langle \log \mbh\rangle$(\oiii)        &$7.24\pm0.08$  &$ 7.31\pm0.16$  & 84.9\%         & isotropic  \\
$\langle M_{\rm bulge}\rangle$           &$-19.42\pm0.11$&$-18.97\pm0.19$ & 11.6\%         & isotropic  \\
$\langle \log {\cal E}\rangle^{a}$       &$-0.98\pm0.05$ &$...$           &   ...          & isotropic  \\
$\langle \log {\cal E}\rangle^{b}$       &$-0.29\pm0.09$ &$-0.49\pm0.14$  & 9.7\%          & isotropic  \\
$\langle \log L_{\rm [O~III]}\rangle$    &$41.57\pm0.06$ &$41.67\pm0.12$  & 16.7\%         & isotropic  \\
$\langle \Gamma_{\rm HX} \rangle$        &$1.76\pm0.03$  &$1.51\pm0.09$   & 0.03\%         & isotropic  \\
\enddata					        	        	
\tablecomments {$ ^{a}$ based on 5100 $\rm \AA$~ luminosity, $^{b}$ based on \oiii luminosity.}
\end{deluxetable} 
\normalsize

\clearpage

\begin{deluxetable}{lccccc}
\tabletypesize{\footnotesize}
\tablewidth{0pt}
\tablecaption{{\sc A Summary of the Statistical Properties} 
\label{tbl-1}}
\tablehead{
Parameters                            & non-HBLR S2s  &    HBLR S2s    & HBLR S2s     & non-HBLR S2s \\
                                      &   (absorbed)  &   (absorbed)   &(unabsorbed)  &(unabsorbed)  } 
\startdata                                                                                                               
$\langle \log \mbh\rangle$(\oiii)     &$6.51\pm0.19$  &$ 7.31\pm0.16$  &$7.34\pm0.23$  &$7.80\pm0.14$  \\
$\langle \log {\cal E}\rangle^{a}$    &$0.23\pm0.14$  &$-0.49\pm0.14$  &$-0.66\pm0.17$ &$-2.76\pm0.27$ \\ \hline
                                      &   NLS1s       &    BLS1s       &               &               \\ \hline
$\langle \log \mbh\rangle$(H$\beta$)  &$ 6.53\pm0.06$ &$7.53\pm0.05$   &    ...        &    ...        \\
$\langle \log \mbh\rangle$(\oiii)     &$ 6.73\pm0.11$ &$7.24\pm0.08$   &    ...        &    ...        \\
$\langle \log {\cal E}\rangle^{a}$    &$-0.16\pm0.05$ &$-0.98\pm0.05$  &    ...        &    ...        \\
$\langle \log {\cal E}\rangle^{b}$    &$ 0.19\pm0.14$ &$-0.29\pm0.09$  &    ...        &    ...        \\ 
\enddata
\tablecomments {$ ^{a}$ based on 5100 $\rm \AA$~ luminosity, $^{b}$ based on \oiii luminosity.}  
\end{deluxetable}
\normalsize


\def\bhmm{$\sim 10^6$}
\def\bhmmm{$\sim 10^7$}
\def\bhmmmm{$\sim 10^8$}

\begin{deluxetable}{lcccccccccc}
\tabletypesize{\footnotesize}
\tablewidth{0pt}
\tablecaption{{\sc Summary of the Seyfert Galaxy Family}
\label{tbl-1}}
\tablehead{
                & \multicolumn{2}{c}{Seyfert 1s}& &\multicolumn{2}{c}{HBLR S2s}& & \multicolumn{3}{c}{non-HBLR S2s}\\\cline{2-3}\cline{5-6}\cline{8-10}
Parameter       & BLS1s    & NLS1               & &absorbed    & unabsorbed    & & absorbed & unabsorbed A& unabsorbed B }
\startdata
Torus orientation$^a$& face-on  &face-on  & & edge-on    & edge-on    &  & edge-on &edge-on & $?$          \\
BH mass ($\sunm$)    &\bhmmm    &\bhmm    & &\bhmmm      &\bhmmm      &  & \bhmm   &\bhmm   & \bhmmmm      \\
Eddington ratio      &$\sim 0.1$&$\sim 1$ & & $\sim 0.1$ & $\sim 0.1$ &  &$\sim 1$ &$\sim 1$&$\le 10^{-3}$ \\ 
Gas-dust ratio       & $ ?$     &$?$      & & high       & low        &  & high    &low     & $?$          \\
     
\enddata
\vskip 0.1cm
\parbox{5.6in}
{The question mark means that this parameter is uncertain. The unabsorbed non-HBLR S2 B is predicted, which is 
characterized by less massive black holes and high accretion rates.} 
\tablecomments {$ ^{a}$ "face-on" refers to there is no obscuration on the line of observer's sight whereas "edge-on"
does there is obscuration.}
\end{deluxetable} 

\end{document}